\documentclass[footinbib,aps,pra,reprint,superscriptaddress]{revtex4-1}

\usepackage{amsmath,amssymb,braket,upgreek,bm,verbatim}
\usepackage{graphicx,tabularx} 
\usepackage{xcolor}

\usepackage[absolute]{textpos}
\usepackage{xcolor}
\usepackage{multirow}
\definecolor{darkblue}{rgb}{0,0,0.5}

\usepackage{hyperref}
\hypersetup{
colorlinks=true,
linkcolor=black,
filecolor=blue,
citecolor=darkblue,  
urlcolor=black,
}

\def\bx{\mathbf{x}}
\def\by{\mathbf{y}}

\def\bbD{{\bm{\mathcal{D}}}}

\def\cF{\mathcal{F}}

\def\Pr{\mathrm{Pr}}

\DeclareMathOperator{\Tr}{Tr}

\newcommand{\imag}{\mathrm{i}}
\begin{document}
\title{Bayesian tomography of high-dimensional on-chip biphoton frequency combs with randomized measurements}

\author{Hsuan-Hao Lu}
\thanks{These authors contributed equally to this work.}
\author{Karthik V. Myilswamy}
\thanks{These authors contributed equally to this work.}
\affiliation{School of Electrical and Computer Engineering and Purdue Quantum Science and Engineering Institute, Purdue University, West Lafayette, Indiana 47907, USA}
\author{Ryan S. Bennink}
\affiliation{Quantum Computational Science Group, Oak Ridge National Laboratory, Oak Ridge, Tennessee 37831, USA}
\author{Suparna Seshadri}
\affiliation{School of Electrical and Computer Engineering and Purdue Quantum Science and Engineering Institute, Purdue University, West Lafayette, Indiana 47907, USA}
\author{Mohammed S. Alshaykh}
\affiliation{Electrical Engineering Department, King Saud University, Riyadh 11421, Saudi Arabia}
\author{Junqiu Liu}
\author{Tobias J. Kippenberg}
\affiliation{Institute of Physics, Swiss Federal Institute of Technology Lausanne (EPFL), 1015 Lausanne, Switzerland}
\author{Daniel E. Leaird}
\author{Andrew M. Weiner}
\affiliation{School of Electrical and Computer Engineering and Purdue Quantum Science and Engineering Institute, Purdue University, West Lafayette, Indiana 47907, USA}
\author{Joseph M. Lukens}
\email{lukensjm@ornl.gov}
\affiliation{Quantum Information Science Section, Oak Ridge National Laboratory, Oak Ridge, Tennessee 37831, USA}

\date{\today}

\maketitle

\begin{textblock}{13.3}(1.4,15)
\noindent\fontsize{7}{7}\selectfont \textcolor{black!30}{This manuscript has been co-authored by UT-Battelle, LLC, under contract DE-AC05-00OR22725 with the US Department of Energy (DOE). The US government retains and the publisher, by accepting the article for publication, acknowledges that the US government retains a nonexclusive, paid-up, irrevocable, worldwide license to publish or reproduce the published form of this manuscript, or allow others to do so, for US government purposes. DOE will provide public access to these results of federally sponsored research in accordance with the DOE Public Access Plan (http://energy.gov/downloads/doe-public-access-plan).}
\end{textblock}

\textbf{Owing in large part to the advent of integrated biphoton frequency combs (BFCs)~\cite{Kues2019}, recent years have witnessed increased attention to quantum information processing in the frequency domain for its inherent high dimensionality and entanglement compatible with fiber-optic networks. 
Quantum state tomography (QST) of such states, however, has required complex and precise engineering of active frequency mixing operations
~\cite{Kues2017, Imany2018, Lu2018b}, 
which are difficult to scale.
To address these limitations, we propose a novel solution that employs 
a pulse shaper and electro-optic phase modulator (EOM) to perform random operations instead of mixing in a prescribed manner. 
We successfully verify the entanglement and reconstruct the full density matrix of BFCs generated from an on-chip Si$_{3}$N$_{4}$ microring resonator (MRR) in up to an $8\times 8$-dimensional two-qudit Hilbert space, 
the highest dimension to date for frequency bins. More generally, our employed Bayesian statistical model~\cite{Blume2010, Lukens2020b} can be tailored to a variety of quantum systems with restricted measurement capabilities, forming an opportunistic tomographic framework that utilizes all available data in an optimal way.
} 

Encoding $d$ levels of quantum information on single photons, known as photonic qudits~\cite{Erhard2020}, offers crucial advantages for quantum communication and networking applications~\cite{Cozzolino2019}, such as higher information capacities~\cite{Barreiro2008}, increased noise tolerance~\cite{Cerf2002, Ecker2019}, and stronger violations of Bell's inequalities~\cite{vertesi2010closing}. Generation and manipulation of photonic qudits have been explored in many degrees of freedom, including path~\cite{Wang2018,Qiang2018}, orbital angular momentum~\cite{Bavaresco2018,Brandt2020}, frequency bins~\cite{Kues2017,Imany2018,Lu2020}, and time bins~\cite{Martin2017,Ikuta2017}. 
Integrated photonics plays a pivotal role in scaling the complexity of quantum states
~\cite{Moody2020,Yang2021} and quantum operations~\cite{Elshaari2020}, and the frequency degree of freedom is particularly attractive as on-chip BFCs can produce a large number of spectrally entangled bins in a compact fashion.

Joint spectral intensity (JSI) measurements are commonly used to characterize BFCs, but such measurements 
are insensitive to phase coherence (and hence entanglement) 
across frequency-bin pairs. Thus, reconstruction of BFC density matrices has been realized through active mixing of frequency bins~\cite{Kues2017, Imany2018, Lu2018b}, 
such that measurements in multiple bases can be realized. In one method, 
by properly setting the amplitude and phase on a pulse shaper and the modulation voltage on a subsequent EOM, one can filter out overlapping sidebands 
to perform projective frequency-bin measurements~\cite{Kues2017, Imany2018}. Alternatively, 
a quantum frequency processor~\cite{Lu2019c}  
can be used to synthesize full quantum gates for tomography ~\cite{Lu2018b}. Nevertheless, both methods face roadblocks en route to higher dimensions: 
aggressive amplitude filtering of the input state is inevitable in the first approach, 
while the number of elements required for arbitrary frequency qudit operations~\cite{Lukens2020a} limits the maximum dimensionality possible with current technology.

\begin{figure*}[tb!]
	\centering
	\includegraphics[trim=0 290 0 0,clip,width=7in]{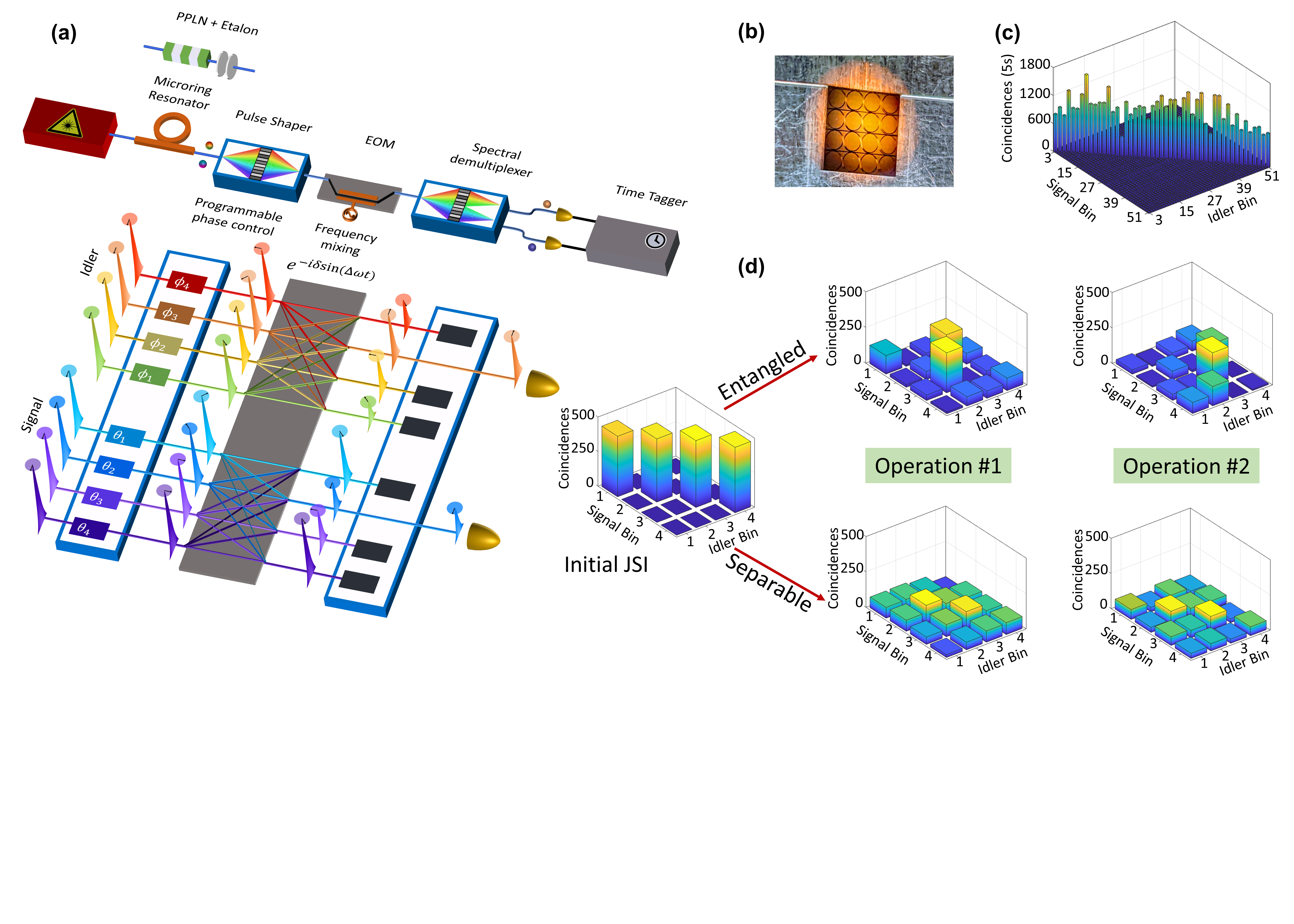}
	\caption{(a) Experimental setup and conceptual illustration of the proposed method. A random operation is uniquely determined by the $d$ random spectral phases $\theta_m$ ($\phi_m$) on the signal (idler) bins and the EOM modulation index $\delta$. (b) Microscope image of the MRR chip. The device in the last column of the second row is used experimentally. (c)~JSI of the output generated by the MRR. (d)~Examples of JSI measurements simulated for the cases of both an entangled state and a classically correlated separable state for two different operations with $\delta=1.5$ rad (left) and $\delta=2$ rad (right) and randomly chosen spectral phases.}
	\label{fig1}
\end{figure*}

Accordingly, these existing methods are ill-suited to single-frequency electro-optic modulation: for them, the infinite Fourier series of Bessel functions produced by sinewave electro-optic modulation---a far cry from standard quantum bases---present a challenge to be overcome. In this work, however, we instead leverage the complex mixing behavior of an EOM to our advantage as built-in randomized measurements for BFC characterization. 
By applying Bayesian tomographic techniques~\cite{Blume2010, Lukens2020b}, we obtain complete state estimates for any dataset, including uncertainties commensurate with the data gathered, 
thus bolstering all results obtained from our novel measurement technique with a principled foundation. Importantly, these Bayesian features extend beyond the specific nuances of frequency-bin encoding to any quantum system, offering the promise of meaningful inference irrespective of whatever experimental constraints may have limited the measurements performed.

Figure~\ref{fig1}(a) illustrates the experimental setup and concept behind our proposed scheme. The states of interest are BFCs 
with mode spacing $\Delta \omega/2\pi \sim40$~GHz and dimension $d$ in both signal and idler photons. The first test source is prepared by pumping a periodically poled lithium niobate (PPLN) waveguide with a continuous-wave laser operating at $\sim$780~nm, followed by filtering the broadband spontaneous parametric down-conversion spectrum with a Fabry-Perot etalon~\cite{Lu2018b,Lu2019c}. The second source exploits spontaneous four-wave mixing in an on-chip Si$_{3}$N$_{4}$ MRR~\cite{liu2020photonic}, where we pump the ring with a tunable continuous-wave laser operating in the optical C-band at one of the ring resonances~\cite{Kues2017,Imany2018}. Ideal maximally entangled states are of the form $\ket{\Psi_d} = \frac{1}{\sqrt{d}} \sum_{m=1}^{d} e^{\imag\alpha_m} \ket{m,m}$, where $\ket{m,m}$ represents the photon pair which is centered at frequency $\omega_0 \pm (m+B)\Delta \omega$ for the signal (idler); $\alpha_m$ is the phase of each pair. The integer $B$ here denotes the number of signal (idler) bins at the center of the biphoton spectrum that are blocked by bandstop filters [omitted in Fig.~\ref{fig1}(a)]. 
Details regarding BFC state preparation can be found in the Methods.

The generated state is then directed to a pulse shaper and an EOM for the implementation of $R_\mathrm{tot}$ randomly chosen operations. For each operation, we apply a set of $d$ random spectral phases $\theta_m$ ($\phi_m$) onto the signal (idler) bins with the pulse shaper. The spectral phases ($\theta_m,\phi_m$) are uniformly sampled between $0$ and $2\pi$. 
The EOM is driven by a sinusoidal voltage with amplitude $\delta$ and frequency equal to the mode spacing $\Delta \omega$, imposing a temporal phase $\exp[-\imag\delta\sin\Delta\omega t]$ onto each photon---equivalently introducing coupling between distinct frequency bins with weights given by Bessel functions of the first kind $J_n(\delta)$. The strength of the imposed phase modulation $\delta$ is selected from a set of $R_\mathrm{tot}$ values equispaced between $\delta\in[0,\delta_{\max}]$, with $\delta_{\max}$ set by the maximum radio-frequency (RF) power attainable at $\Delta\omega$. 
Among these $R_\mathrm{tot}$ operations, we designate $\delta=0$ for the first measurement (i.e., the EOM turned off, making it a conventional JSI), while the modulation indices for the remaining $R_\mathrm{tot}-1$ operations are chosen in a random order from the equispaced set, without replacement. 

\begin{figure*}[tb!]
	\centering
	\includegraphics[trim=0 330 0 0,clip,width=7in]{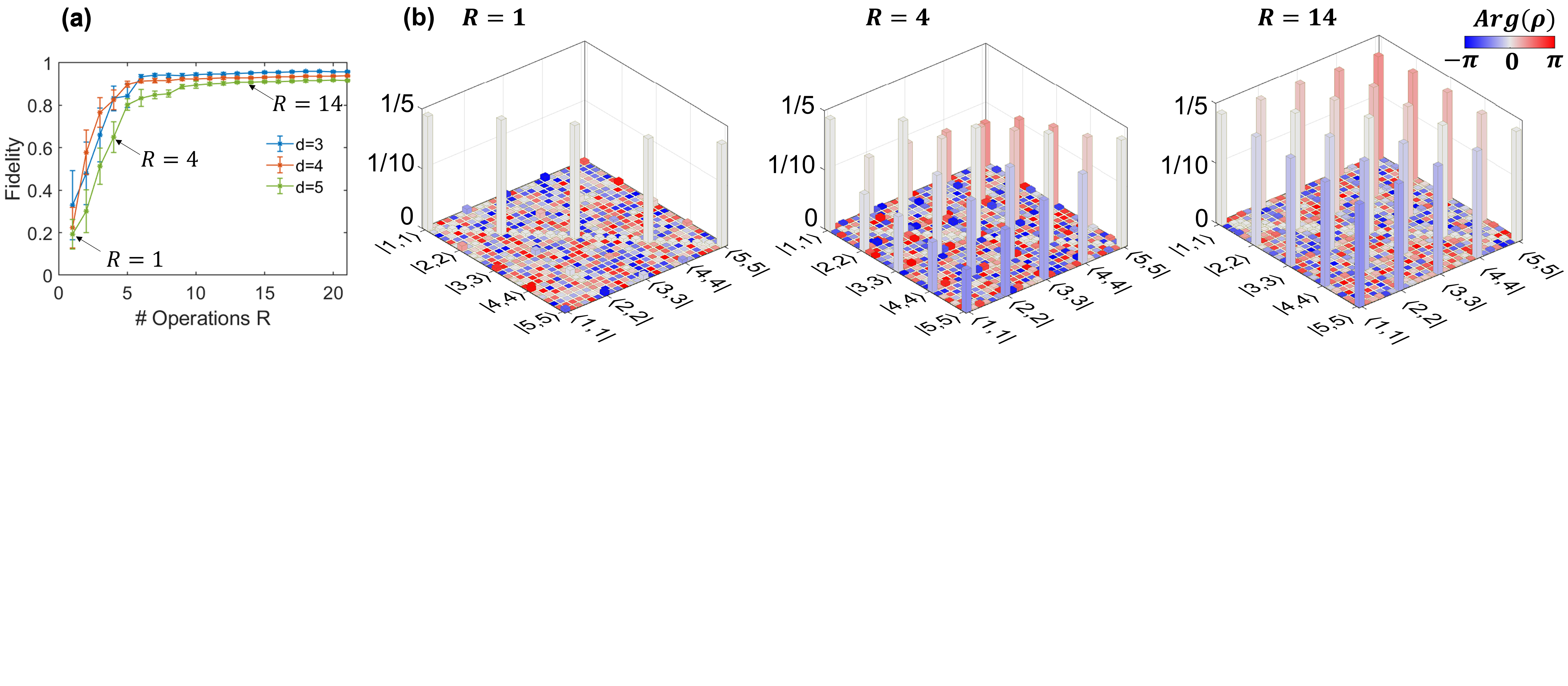}
	\caption{Evolution of the estimated state for the PPLN BFC. (a) Fidelity with respect to a maximally entangled state as a function of random operations used in Bayesian inference for $d\in \{3,4,5\}$. (b) Retrieved density matrices for $d=5$ when the first $R\in\{1,4,14\}$ random operations are considered; their corresponding fidelities are marked by arrows in (a). Shading indicates the phase according to the colorbar scheme shown (here and in the following figures).}
	\label{fig2}
\end{figure*}

The photons are then passed to a 
wavelength-selective switch and we scan the filters to collect coincidences over the original $d\times d$ frequency mode grid for each implemented operation, 
omitting any photons scattered outside of this computational space. Figure~\ref{fig1}(d) shows examples  of such measurements, simulated using multinomial statistics, corresponding to two different random operations for the cases of a classically correlated separable state ($\rho_\mathrm{sep} = \frac{1}{d}\sum_{m=1}^d \ket{m,m} \bra{m,m}$) and a maximally entangled quantum state($\rho_\mathrm{ent} = \frac{1}{d}\sum_{m,n=1}^d \ket{m,m} \bra{n,n}$) for $d=4$. In the absence of modulation ($\delta=0$), their JSIs are identical, 
yet when the EOM is turned on, the frequency correlations vary strongly. For example, in the extreme case of complete incoherence between energy-matched bins ($\rho_\mathrm{sep}$ here), applying random phases on the initial pulse shaper has no impact on the measured output. Such differences imply that a collection of these measurements can be exploited to infer the full density matrix.

Given our knowledge of the quantum operations applied and a set of $R$ $d\times d$ coincidence results, we then employ Bayesian tomographic techniques~\cite{Blume2010, Lukens2020b} to estimate the input quantum state. 
Conceptually simple---though numerically challenging---Bayesian QST assigns a posterior probability distribution to all unknown parameters $\bx$, given a set of observations $\bbD$, according to Bayes' theorem, $\Pr(\bx|\bbD) = \Pr(\bbD|\bx) \Pr(\bx)/\Pr(\bbD)$, which incorporates both a physical model through $\Pr(\bbD|\bx)$ and any prior information in $\Pr(\bx)$. Significantly, the estimator formed by averaging any quantity of interest over the posterior $\Pr(\bx|\bbD)$ is guaranteed to offer the lowest squared error on average~\cite{Robert1999}, making the Bayesian mean provably optimal for any number of measurements. 
Using the model and algorithm described in the Methods, we obtain estimates 
of the density matrix $\rho$, fidelity $\mathcal{F}_d$, and logarithmic negativity $E_d$; $E_d>0$ is a sufficient condition for nonseparability, and $E_d$ upper bounds distillable entanglement~\cite{Peres1996,Vidal2002}.

\begin{figure*}[tb!]
	\centering
	\includegraphics[trim=0 0 0 0,clip,width=7in]{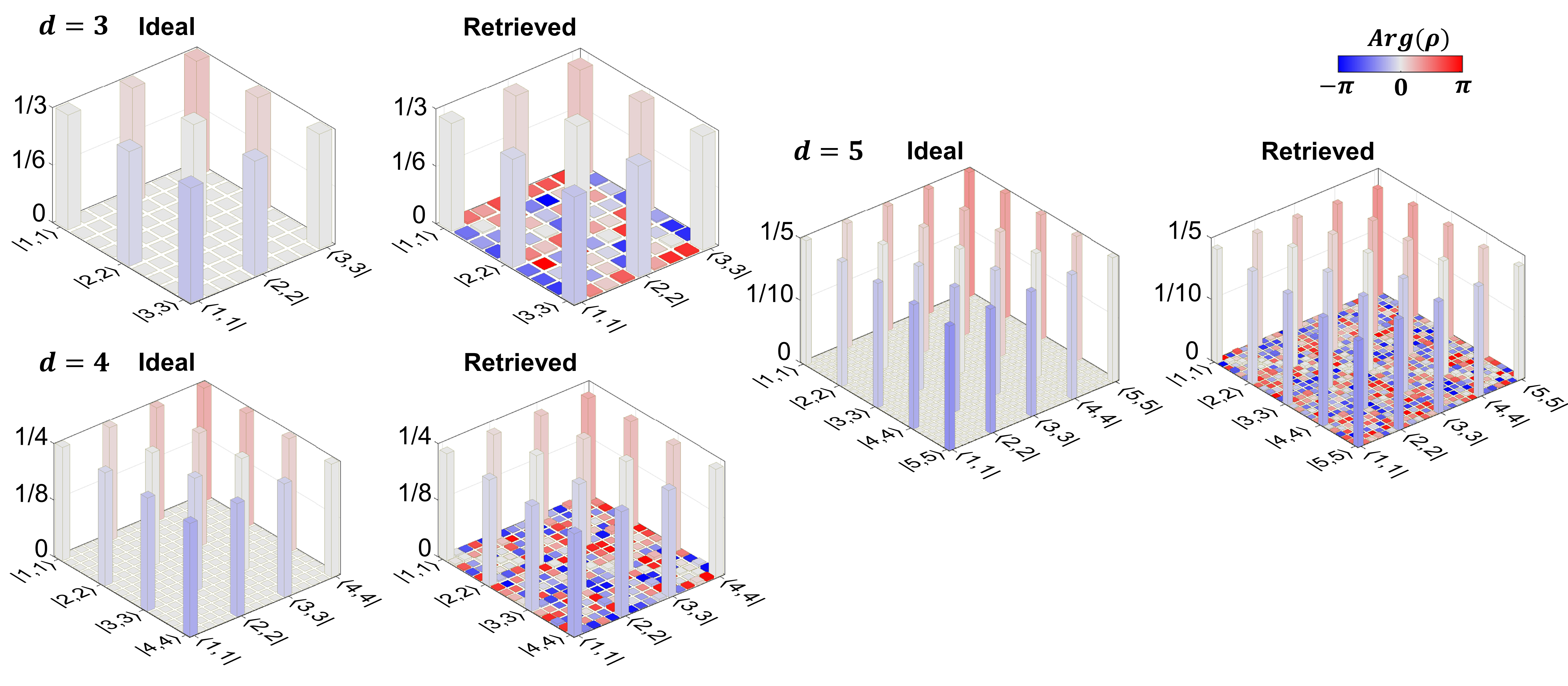}
	\caption{Final estimated states for the PPLN BFC and $R=21$ measurements. Ideal and retrieved density matrices for $d\in\{3,4,5\}$. Ideal states are $d$-dimensional Bell states with additional spectral phase equivalent to 20~m of single-mode fiber.}
	\label{fig3}
\end{figure*}

For PPLN experiments, we consider BFCs of qudit dimension $d\in\{3,4,5\}$. 
We implement a total of $R_\mathrm{tot}=21$ randomly chosen operations for each dimension, with a maximum modulation index $\delta_{\max}= 2.5$~rad. 
We compute the fidelity of the retrieved density matrices at various stages of Bayesian estimation with respect to the ideal state $\ket{\Psi_d}$ with $\alpha_m = \beta_2 L \Delta\omega^2 (m+B)^2$, corresponding to dispersion accumulated over $L=20$~m of single-mode fiber (approximate length of fiber between the PPLN source and EOM). Figure~\ref{fig2}(a) shows the evolution of the Bayesian-estimated fidelity after $R\leq R_\mathrm{tot}$ random operations are performed. (For these and all results in this paper, no subtraction of accidentals is performed.) In Fig.~\ref{fig2}(b), we plot the mean density matrices retrieved from Bayesian analysis for $d=5$, at specific numbers of random operations. For the case of $R=1$, the density matrix resembles a separable state with small off-diagonal elements. 
As $R$ increases, the off-diagonal elements rise 
as the phase coherence in the BFC is revealed by operations involving frequency mixing; the fidelity with respect to the ideal state increases accordingly, converging at $R\approx 10$ operations for all three dimensions (see Methods for discussion of possible explanations of this convergence behavior).

In Fig.~\ref{fig3}, we plot both the ideal and the final estimated density matrices ($R=R_\mathrm{tot}=21$). Both absolute values and phases align well with theory---the only discrepancy being the inconsequential phase values of the near-zero off-diagonal elements. From the Bayesian results, we report fidelities $\mathcal{F}_d$ of $\mathcal{F}_3=(95.8\pm0.4)\%$, $\mathcal{F}_4=(94.0\pm0.4)\%$, and $\mathcal{F}_5=(91.7\pm0.4)\%$, 
which are in the neighborhood of the theoretically predicted $\mathcal{F}_3=97.1\%$, $\mathcal{F}_4=96.0\%$, and $\mathcal{F}_5=94.9\%$ for a white noise model $\rho_{\lambda} = \lambda\ket{\Psi_d}\bra{\Psi_d} + \frac{1-\lambda}{d^2} I_{d^2}$, with $\lambda$ chosen such that the coincidences-to-accidentals ratio ($\mathrm{CAR}$) matches the experimentally measured value of 90 (see Methods). The Bayesian-estimated log-negativities are $E_3=1.523\pm0.006$ ebits, $E_4=1.911\pm0.006$ ebits, and $E_5=2.198\pm0.007$ ebits, comparable with those of $\ket{\Psi_d}$ ($\log_2 d$) given by 1.58, 2, and 2.32 ebits, respectively.


\begin{figure*}[tb]
	\centering
	\includegraphics[trim=0 0 0 0,clip,width=7in]{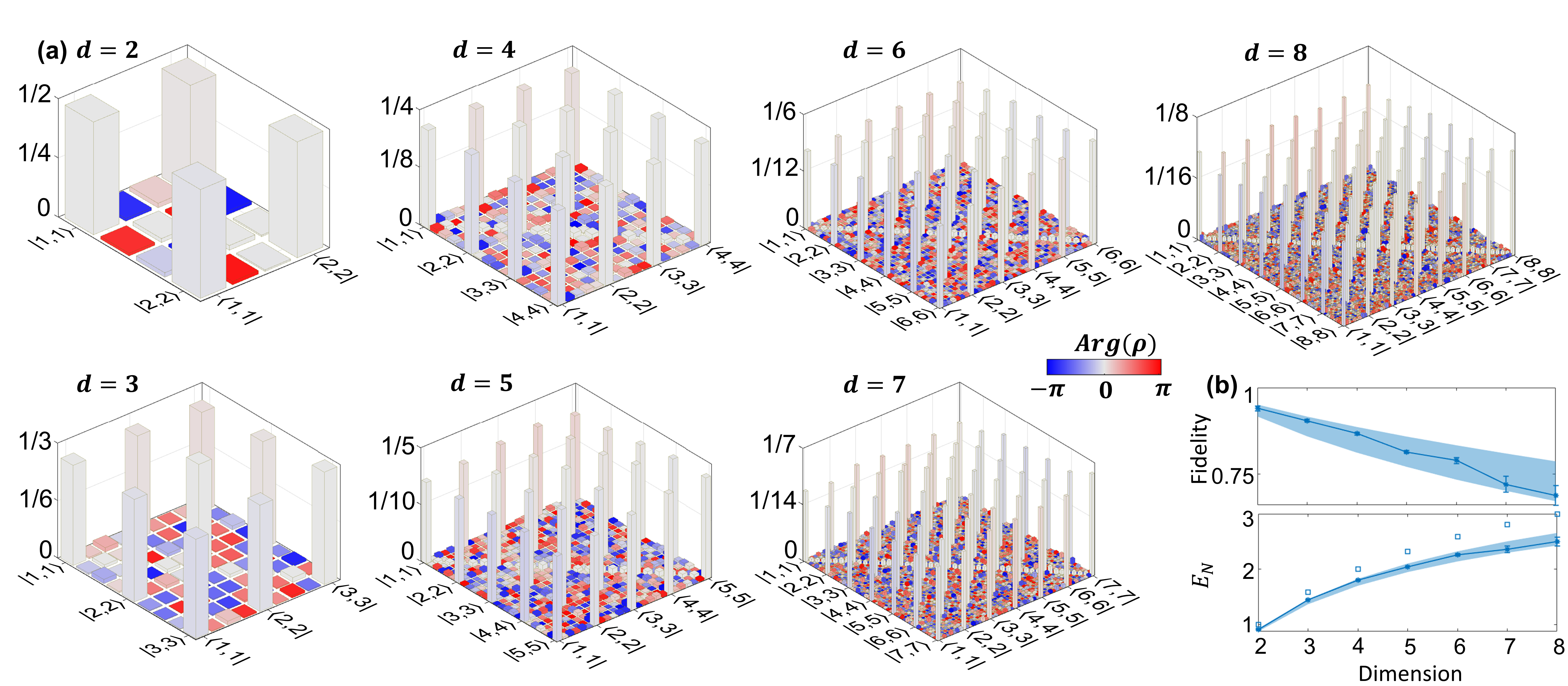}
	\caption{Tomography results for MRR BFCs and $R=30$ measurements. (a) Retrieved mean density matrices for dimensions up to $d=8$. Ideal states $\ket{\Psi_d}$ (not plotted) are $d$-dimensional maximally entangled states with uniform phase. (b) Fidelities and log-negativities of retrieved density matrices (solid line). Shaded region corresponds to theoretically computed values for $\rho_{\lambda}$ (a mixture of a maximally entangled state and white noise) lower- and upper-bounded by CARs of $17$ and $30$, respectively. Square markers correspond to log-negativities of maximally entangled states, given by $\log_2 d$.}
	\label{fig4}
\end{figure*}

We then adopt the same methodology for BFCs generated using an on-chip Si$_{3}$N$_{4}$ MRR [Fig.~\ref{fig1}(b)]\cite{liu2020photonic}. 
The measured JSI is shown in Fig.~\ref{fig1}(c), with coincidences recorded over $49$ signal-idler bin pairs, eight of which (bins 23--30) are selected for testing. The estimated on-chip pair generation rate varies between $\sim 1.3\times 10^{6}$ and $\sim 2.2\times 10^{6}$ s$^{-1}$ per frequency-bin pair, 
and the CAR---here defined as the ratio of a given diagonal element to the average of all off-diagonal elements---lies in the interval $[17,30]$ for the eight bins we test. We perform tomography for qudit dimension $d\in\{2,3,\hdots,8\}$ with $R_\mathrm{tot}=30$ operations for each $d$ and $\delta_{\max}=3.4$~rad (increased from $2.5$~rad in the PPLN case due to reoptimization of the RF setup for lower loss). 
We also apply additional spectral phases that compensate for the residual biphoton phase (see Methods), 
and thus we compute the fidelities of retrieved density matrices with respect to the ideal state $\ket{\Psi_d}$ with $\alpha_m = 0$. 

In Fig.~\ref{fig4}(a), we plot the final estimated density matrices. 
The elements indexed by $\ket{m,m}\bra{n,n}$ have strong nonzero amplitudes, agreeing well with theory for ideal entangled states.  The background corresponding to energy-mismatched bins (gray baseline on the diagonal) is consistent with a white noise model and real-valued owing to hermiticity. 
Figure~\ref{fig4}(b) plots the fidelities with respect to $\ket{\Psi_d}$  and the corresponding log-negativities $E_d$, 
lying comfortably within the range predicted for our noise model using experimentally observed CARs (shaded region). 
Significantly, our $d=8$ result of $E_8=2.50 \pm 0.08$ ebits can only be achieved by two-qudit states with $d\geq 6$, indicating the genuine high-dimensional nature of the observed entanglement. These results showcase the highest dimension of a fully reconstructed density matrix---Hilbert space dimension of 64---in experimental frequency-bin encoding. 

Technologically speaking, our approach aligns closely with recent ideas presented in EOM-based frequency-bin quantum random walks~\cite{Imany2020}, where here we precede the walk with random spectral phases and consider varying circuit depths. Yet beyond the confines of frequency-bin encoding, our statistical treatment hints at the much wider value of Bayesian models in quantum information.
Neither the choice of measurement settings nor number of datapoints has any bearing on the legitimacy of Bayesian tomography~\cite{Blume2010, Lukens2020b}; hence, Bayesian estimation will return a reasonable result for any dataset, with automatic uncertainty quantification indicating the confidence warranted from the data. 
This feature imparts Bayesian inference with unique flexibility compared to other advanced random measurement approaches, in that it does not assume, e.g., low-rank states~\cite{Gross2010} or rely on unitary operations drawn from specific distributions~\cite{Brydges2019, Huang2020}. Therefore, any quantum system for which an appropriate physical model can be constructed is ripe to potentially benefit from Bayesian models like the one presented here.

Although not required conceptually, well-chosen measurements are \emph{practically} valuable for obtaining final estimates accompanied by low uncertainties. 
Our experimental results show through example that the  datasets from random modulation are more than sufficient to converge from an initially uniform (Bures) prior to density matrices with small error bars and in good agreement with the expected ground truth. Indeed, arguments from a simple theoretical model suggest that random EOM measurements with $\delta_\mathrm{max} \sim \mathcal{O}(d)$ efficiently cover the entire Hilbert space of a $d$-dimensional quantum system (see Methods for details),
so that our measurement approach offers a straightforward path for high-dimensional frequency-bin characterization and 
should open new opportunities for BFCs in quantum information processing.

\section*{Acknowledgments}
We thank AdvR for loaning the PPLN ridge waveguide; P. Imany, N.~B. Lingaraju, and A.~J. Moore for valuable discussions; A.~A.~N. Ovi for laboratory help; and B.~T. Kirby for introducing us to the Bures distribution. This work was performed in part at Oak Ridge National Laboratory, operated by UT-Battelle for the U.S. Department of Energy under contract no. DE-AC05-00OR22725. Funding was provided by the U.S. Department of Energy, Office of Advanced Scientific Computing Research, Early Career Research Program (Field Work Proposal ERKJ353), the National Science Foundation (1839191-ECCS, 2034019-ECCS), the Air Force Office of Scientific Research (Award no. FA9550-19-1-0250), and the Swiss National Science Foundation under Grant no. 176563 (BRIDGE). K.V.M. acknowledges support from the QISE-NET fellowship program of the National Science Foundation (DMR-1747426). M.S.A. acknowledges support from the College of Engineering Research Center at King Saud University. The Si$_{3}$N$_{4}$ samples were fabricated in the EPFL Center of MicroNanoTechnology (CMi).

\section*{Author contributions}
H.H.L. and K.V.M. contributed equally to this work. H.H.L. initiated the concepts, led the experiments, and contributed to the theoretical analysis. K.V.M. performed experiments with MRR sources and led writing of the paper. S.S., M.S.A, and D.E.L assisted the experiments. R.S.B. contributed to the theoretical analysis. J.L. designed and fabricated the integrated device, supervised by T.J.K. A.M.W. supervised the work at Purdue and assisted with system design and analysis. J.M.L. managed the theoretical analysis and numerical simulation, and supervised the project. All authors reviewed the results and contributed to the manuscript.


\section{Methods}
\label{methods}

\subsection{PPLN source setup}

Our test source is a 2.1~cm-long fiber-pigtailed PPLN ridge waveguide (AdvR), possessing an internal efficiency of 150 $\%/$W for second-harmonic generation and fiber-coupling efficiency of $\sim$75\% per facet. We couple a continuous-wave laser (Toptica) operated at $\sim$5~mW and $\sim$780~nm into the PPLN waveguide, temperature-controlled to 51.6 $^{\circ}$C for SPDC under type-0 phase matching. Broadband spectrally entangled photon pairs spanning $>$5 THz are generated and subsequently filtered by a tunable fiber-pigtailed Fabry-Perot etalon (Luna Innovations) with 20~GHz mode spacing and a full-width at half-maximum linewidth of 1 GHz. We carefully tune the etalon's temperature to align its transmission peaks with the generated entangled photons, i.e., to maximize the coincidences between the symmetric, spectrally filtered mode pairs. 

To minimize crosstalk in our 20~GHz-resolution demultiplexer (Finisar Waveshaper 4000S/X), we utilize the first pulse shaper (Finisar Waveshaper 1000S) to perform amplitude filtering (on top of the phase masks programmed for the QST) and block every other frequency bin, resulting in a 40~GHz-spaced BFC with a measured CAR$\approx$90. A 300~GHz bandstop filter (corresponding to the case of $B=3$) is programmed on the same pulse shaper in the middle of the BFC spectrum, which allows us to apply strong modulation on the EOM without the possibility of the signal photon jumping over into the idler's modes, and vice versa. For coincidence measurements, we use a window of 128~ps and integration times of 20~s for all dimensions. 

\begin{figure}
\centering
	\includegraphics[trim=25 350 800 0,clip,width=3.5in]{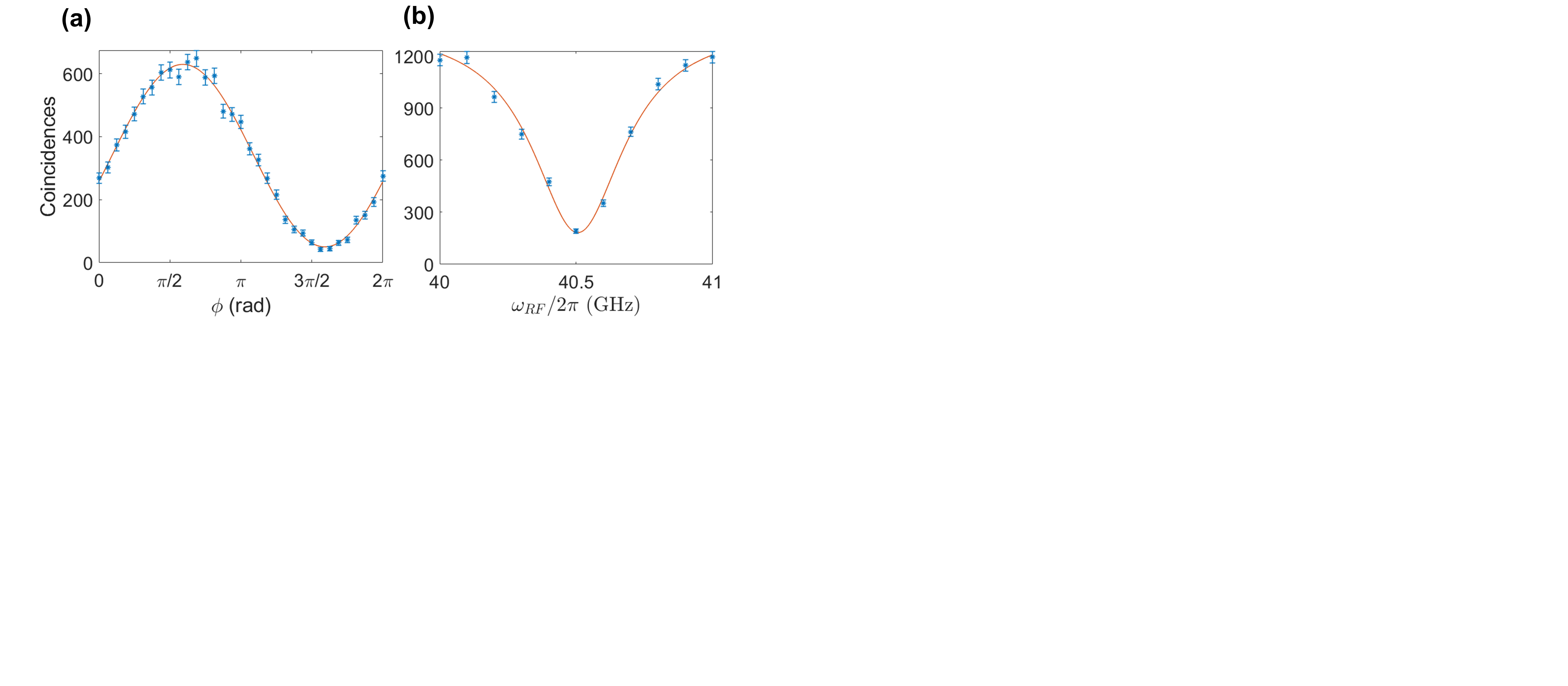}
	\caption{Characterization of residual BFC spectral phase and FSR of MRR. Two adjacent signal-idler bin pairs are selected and equally mixed using an EOM. Biphoton time-correlation functions of one energy-matched pair are then measured. (a)~Coincidences at $\tau=0$ as joint spectral phase $\phi$ is swept. (b)~Coincidences integrated over $\tau$ as modulation frequency $\omega_{RF}$ is swept at a spectral phase that minimizes coincidences at $\tau=0$ (details in text).}
	\label{fig5}
\end{figure}

\subsection{MRR source set up}
The Si$_{3}$N$_{4}$ MRR used in our experiment is fabricated using the optimized photonic Damascene reflow process \cite{liu2021high} with a cross-section of 2~$\upmu$m$\times$ 0.95~$\upmu$m. Such a process has enabled ultralow loss waveguides that have paved the way for dissipative Kerr solitons with free spectral ranges (FSRs) as low as 10~GHz~\cite{liu2020photonic}. In the current work, the radius of the ring is 561~$\upmu$m, corresponding to an FSR of 40.5~GHz---within the range of commercial EOMs---allowing us to drive the EOM at a frequency equal to the FSR for random operations.  The gap between the ring and the bus waveguide is 0.3~$\upmu$m, resulting in strong overcoupling with an intrinsic $Q$-factor of $\sim$10$^7$ and a loaded $Q$-factor of $\sim$10$^6$. 
We pump the ring using a tunable continuous-wave laser (New Focus) operating in the optical C-band, with an on-chip power of $\sim$10~mW---well below the classical comb threshold of $\sim$80~mW. The pump is amplified using an erbium doped fiber amplifier (EDFA), which is subsequently filtered using a set of two 100~GHz-wide dense wavelength division multiplexing (DWDM) filters 
to suppress amplified spontaneous emission from the EDFA. Lensed fibers are used to couple the pump into the MRR, which is positioned on a temperature-controlled stage maintained at $\sim$23 $^{\circ}$C. The fiber-to-fiber coupling loss of the ring is around $\sim$4~dB. Such a low loss was realized using engineered inverse waveguide tapers \cite{liu2018double}. When the pump is tuned into the ring resonance ($\sim$1550.5~nm) and operated below threshold for classical comb generation, it gives rise to BFC states. Since the pump and newly generated biphotons are in the same wavelength band, it is essential to suppress the residual pump after the ring to reduce accidentals in coincidence measurements. We use a set of three 200~GHz-wide DWDM filters, which when combined with bandstop filters in the pulse shaper and demultiplexer 
gives a net pump suppression of $\sim$100~dB. We also tap a portion of the pump power to track its wavelength using a wavelength meter, setting up a computer-based feedback loop to ensure that the pump is operated at the intended resonance. For coincidence measurements, we use a window of 2048~ps (roughly equivalent to the inverse resonance linewidth) and integration time of 5~s for $d\in\{2,3,\hdots,6\}$. For $d\in\{7,8\}$, we reduce the integration time to 3~s in the interest of minimizing the total experimental duration. 

It is necessary to experimentally characterize the FSR of the MRR to determine the precise modulation frequency needed. 
In addition to this, we also characterize the residual spectral phase accumulated by the biphotons, likely due to fiber dispersion, to precompensate for it. For characterization of these quantities, the experimental setup remains the same as shown in Fig.~\ref{fig1}(a). We select two adjacent signal-idler bin pairs in the first pulse shaper and drive the EOM at a frequency $\omega_{RF}$ with amplitude $\delta=1.43$ such that $\left|J_0(\delta)\right|=\left|J_1(\delta)\right|$ for equal mixing. We then pass one energy-matched pair of signal-idler frequency bins through the demultiplexer, now consisting of equal contributions from the adjacent bin due to phase modulation, and measure the biphoton time-correlation function. The theoretical expression for the coincidence rate, assuming both bins have identical Lorentzian lineshape and equal probability amplitude, is given by \cite{KarthikCLEO}
\begin{equation}
\label{eq1}
    R (\tau) \propto e^{-\gamma \left|\tau\right|} \left\{ 1 - \cos\left[\phi+\phi_0-\left(\omega_{FSR}-\omega_{RF}\right)\tau\right]  \right\}
\end{equation}
where $\gamma$ represents the Lorentzian linewidth, $\tau$ the signal-idler delay, $\phi$ the joint spectral phase between the bins applied by the pulse shaper, $\phi_0$ the residual biphoton phase, and $\omega_{FSR}$ the FSR. In Fig.~\ref{fig5}(a), we plot the coincidences at $\tau=0$ as a function of $\phi$ for a set of two adjoining signal-idler bin pairs, where $\tau=0$ is defined as the peak of the histogram in the unmodulated case. The offset of the experimentally obtained cosine function from the origin can be used to deduce the residual spectral phase between the bins. This process is repeated for all sets of adjoining signal-idler bin pairs used in our experiment to deduce the residual spectral phase of the entire biphoton state. For finding $\omega_{FSR}$, we again work with a set of two adjoining signal-idler bin pairs and operate at a spectral phase that minimizes the coincidences at $\tau=0$. In Fig.~\ref{fig5}(b) we plot the coincidences integrated over $\tau$ as we sweep the applied RF frequency $\omega_{RF}$. When $\omega_{RF}=\omega_{FSR}$, the coincidences are minimized, so that 
Fig.~\ref{fig5}(b) implies $\omega_{FSR}/2\pi=40.5$~GHz. The solid lines in Fig.~\ref{fig5} are theoretical estimates, scaled and vertically offset to match the data points via least squares.  The vertical offset and the scaling account for both the nonzero accidentals and unequal bin probability amplitudes. The width of the trace in Fig.~\ref{fig5}(b) can be expressed as $2\gamma$ from Eq.~\eqref{eq1}, from which $\gamma/2\pi\approx200$~MHz is inferred. These obtained resonator linewidth and FSR values are consistent with linear spectroscopy measurements.   

\subsection{Bayesian inference model}
The ultimate goal of the inference process is to estimate the full BFC state, which can be represented as a $d^2\times d^2$ density matrix $\rho$. The frequency bins of the signal (idler) qudit possess annihilation operators $\hat{a}_k^{(S)}$ ($\hat{a}_l^{(I)}$) where $k,l \in \{1,...,d\}$ correspond to center frequencies of $\omega_k = \omega_0 + (k+B)\Delta \omega$ and $\omega_l = \omega_0 - (l+B)\Delta \omega$. 
The density matrix of interest can be formally written as
\begin{equation}
\label{eq:rho}
\rho = \sum_{k,l,k',l'=1}^d \rho_{(kl)(k'l')} [\hat{a}_k^{(S)}]^\dagger [\hat{a}_l^{(I)}]^\dagger \ket{\mathrm{vac}}\bra{\mathrm{vac}}  \hat{a}_{k'}^{(S)} \hat{a}_{l'}^{(I)},
\end{equation}
where $\ket{\mathrm{vac}}$ is the vacuum state. 

We use the index $s$ to denote a specific measurement, which consists of a particular EOM and pulse shaper setting $r(s)\in\{1,…,R_\mathrm{tot}\}$, the signal frequency bin measured $m(s) \in\{1,…,d\}$, and the idler bin measured $n(s) \in\{1,…,d\}$. For notational convenience, the explicit $s$-dependence is suppressed in many of the formulas below, but remains implied for $(r,m,n)$.
For a given $r$, the frequency bins undergo unitary transformations into the output modes via 
\begin{equation}
\label{eq:unit1}
\begin{aligned}
\hat{b}_m^{(S)} = \sum_{k=1}^d V_{mk}^{(r)} \hat{a}_k^{(S)} \\
\hat{b}_n^{(I)} = \sum_{l=1}^d W_{nl}^{(r)} \hat{a}_l^{(I)}.
\end{aligned}
\end{equation}
The unitary operations $V^{(r)}$ and $W^{(r)}$ consist of line-by-line phase shifts on each of the input modes, followed by sinewave electro-optic modulation. 
The index $r$ defines the phase shifts and modulation index for a specific setting: $\theta_n^{(r)}$, $\phi_n^{(r)}$, $\delta^{(r)}$. We can therefore write
\begin{equation}
\label{eq:unit2}
\begin{aligned}
V_{mk}^{(r)} = J_{m-k}(\delta^{(r)}) e^{\imag\theta_k^{(r)}} \\
W_{nl}^{(r)} = J_{l-n}(\delta^{(r)}) e^{\imag\phi_l^{(r)}},
\end{aligned}
\end{equation}
where $J_n(\cdot)$ is the Bessel function of the first kind, and for definiteness we have assumed a modulation function of the form $\exp[-\imag\delta^{(r)}\sin\Delta\omega t]$. 
The only major difference between the signal and idler equations is the input/output index reversal in the Bessel function order, which results from our definition of signal frequencies that increase with index and idler frequencies that decrease with index.

After these operations, we look for coincidences between bins $m$ and $n$, the probability of which can be computed for the input density matrix as
\begin{widetext}
\begin{equation}
\label{eq:prob}
\begin{aligned}
p_s & = \braket{m,n|\rho|m,n} = \braket{\mathrm{vac}|\hat{b}_m^{(S)}\hat{b}_n^{(I)}\rho[\hat{b}_m^{(S)}]^\dagger [\hat{b}_n^{(I)}]^\dagger | \mathrm{vac}} \\
 & = \sum_{k,l,k',l'=1}^d \rho_{(kl)(k'l')}  V_{mk}^{(r)} W_{nl}^{(r)} [V_{mk'}^{(r)}]^* [W_{nl'}^{(r)}]^* \\ & 
 = \sum_{k,l,k',l'=1}^d \rho_{(kl)(k'l')} 
  J_{m-k}(\delta^{(r)})  J_{l-n}(\delta^{(r)})  J_{m-k'}(\delta^{(r)}) J_{l'-n}(\delta^{(r)})
  e^{\imag\left[\theta_k^{(r)} + \phi_l^{(r)} - \theta_{k'}^{(r)} - \phi_{l'}^{(r)}\right]},
\end{aligned}
\end{equation}
\end{widetext}
using Eqs.~(\ref{eq:rho}--\ref{eq:unit2}) to simplify. This expression provides a linear mapping from the density matrix elements $\rho_{(kl)(k'l')}$ to each output probability $p_s$.

We then need to relate these probabilities to the observed coincidence counts $N_s$ through an appropriate likelihood function. In typical quantum tomographic contexts where the full Hilbert space is detected at each setting, a multinomial model is the most conceptually straightforward~\cite{Blume2010}; in our case this would consist of products of the factors $p_s^{N_s}$. However, such a model does not readily apply to situations in which some of the outcomes are unmonitored. In our particular experiment, many bins outside of the original $d^2$-dimensional computational space can be populated, owing to the nonzero Bessel function weights in Eq.~\eqref{eq:prob}. As a rule of thumb, the success probability (i.e., the possibility of a single photon staying in the original $d$-mode computational space) is roughly $1/2$ when the modulation depth $\delta\sim\mathcal{O}(d)$~\cite{Lu2018a}. Rather than attempting to measure all output mode combinations---which in principle involves an infinite-dimensional Hilbert space and in practice means the addition of many measurements with few counts---we focus only on the central $d\times d$ space here. 

On the model side, we can account for unobserved outcomes by introducing an additional flux parameter $K$, defined as the average number of total coincidences that \emph{would} be measured if all $(m,n)$ combinations were tested. Since the EOM and pulse shaper operations are unitary when considered over all modes---apart from an overall insertion loss that does not vary with setting $r$---this scale factor is fixed for all measurement settings. It also automatically accounts for efficiency; explicitly, it can be written as $K=\eta_S \eta_I \Phi \Delta T$, where $\eta_S$ ($\eta_I$) is the total system efficiency from generation through detection for the signal (idler), $\Phi$ the photon pair generation flux, and $\Delta T$ the integration time. Although these quantities themselves could be inferred by considering both singles counts as well as coincidences~\cite{Williams2017, Lu2018b, Lu2019a}, they are not of interest in the present investigation, and so through $K$ we are able to reduce to their combined effect only.

Thus, the mean number of coincidences for a specific setting $s$ becomes $Kp_s$, which we can model with a Poissonian distribution, the product of which produces the full likelihood $L_\bbD(\rho,K) \propto \Pr(\bbD|\rho,K)$
\begin{equation}
\label{eq:LL}
L_\bbD(\rho,K)  = \prod_{s=1}^{Rd^2} e^{-Kp_s} (Kp_s)^{N_s},
\end{equation}
where $\bbD=\{N_1,...,N_{Rd^2}\}$ denotes the set of measured coincidences for all $Rd^2$ settings. This likelihood matches the form we adopted previously in the construction of JSIs (not the full quantum state) from random measurements~\cite{Simmerman2020}. And not only does it readily handle probabilities that do not encompass the full Hilbert space, but it also more accurately reflects the physical situation of high-dimensional measurements with single-outcome detectors. For example, our use of raster scanning with two single-photon detectors means that, for any given configuration, we measure events only for a specific $(m,n)$ frequency-bin pair. Although one can pool the results for all pairs $(m,n)$ and view them synthetically as resulting from a single $\Delta T$ integration time of true $d^2$-outcome measurements---the situation assumed by a multinomial distribution---this ultimately does not align with the actual measurement procedure. Accordingly, the Poissonian likelihood provides both a practical and conceptually satisfying model for our tomographic scenario.

With the likelihood defining the relationship between a given density matrix and the observed data, Bayesian inference next requires specification of a suitable prior distribution for $\rho$ and $K$. In the case of QST, uniform priors are generally preferred, as these apply appreciable weights to all possible states and thereby minimally bias the final results. Although a variety of reasonable uninformative priors can be posited for density matrices, we select the Bures distribution, which enjoys a unique position as the single monotone metric which reduces to both the Fisher and Fubini--Study metrics in the classical and pure state limits, respectively~\cite{Sommers2003}, and in this sense can claim preference as a ``definitive'' uniform prior for Bayesian inference~\cite{Osipov2010}.

To work with the Bures ensemble, it is convenient to express any $d^2\times d^2$ density matrix $\rho$ in the computational basis as
\begin{equation}
\label{eq:Bures}
\rho = \frac{(I_{d^2} + U)G G^\dagger (I_{d^2} + U^\dagger)}{\Tr\left[(I_{d^2} + U)G G^\dagger (I_{d^2} + U^\dagger)\right]},
\end{equation}
where the $d^2\times d^2$ matrices $I_{d^2}$, $U$, and $G$ are the identity, a unitary matrix, and a general complex matrix, respectively. This expression automatically satisfies all physicality conditions (unit-trace, hermiticity, and positive semidefiniteness); by sampling $U$ from the Haar distribution and $G$ from the Ginibre ensemble, the $\rho$ thus formed represents a single draw from the Bures distribution~\cite{Osipov2010}. We can in turn parameterize $\rho$ by the complex vector $\by = (y_1,...,y_{2d^4})$, with each $y_k$ observing a complex standard normal distribution $y_k\overset{\mathrm{i.i.d.}}{\sim} \mathcal{CN}(0,1)$. $d^4$ of the components comprise the $d^2\times d^2$ elements of the Ginibre matrix $G$ directly, while the remaining $d^4$ parameters form a second Ginibre matrix which is converted to the unitary $U$ through the Mezzadri algorithm~\cite{Mezzadri2007}, thereby ensuring Haar randomness.

In addition to the parameters forming $\rho$, the scale factor $K$ must also be suitably parameterized. Following~\cite{Simmerman2020}, we find it convenient to write $K=K_0(1+\sigma z)$, where $K_0$ and $\sigma$ are hyperparameters defined separate of the inference process, and $z$ is taken to follow a standard normal distribution $\mathcal{N}(0,1)$, leading to a normal prior on $K$ of mean $K_0$ and standard deviation $K_0\sigma$. We take $\sigma=0.1$ and $K_0$ equal to the sum of the counts in all $d^2$ bins for the first JSI measurement ($r=1$), where the absence of modulation ensures that all initial photon flux remains in measured bins, i.e., $K_0 = \sum_{s=1}^{d^2} N_s$. This provides an effectively uniform prior, since a fractional deviation of 0.1 is much larger than the maximum amount of fractional uncertainty $1/\sqrt{K_0}\approx 0.02$ expected from statistical noise at our total count numbers; the use of a normal distribution simplifies the sampling process.

The total parameter set can therefore be expressed as the vector $\bx=(\by,z)=(y_1,...,y_{2d^4}, z)$, with the prior distribution
\begin{equation}
\label{eq:prior}
\pi_0(\bx) \propto \left(\prod_{k=1}^{2d^4} e^{-\frac{1}{2}|y_k|^2} \right) e^{-\frac{1}{2}z^2}.
\end{equation}
We note that this parameterization entails a total of $4d^4+1$ independent real numbers ($2d^4$ complex parameters for $\rho$, one real parameter for $K$)---noticeably higher than the minimum of $d^4-1$ required to uniquely describe a density matrix. Nevertheless, this $\rho(\by)$ parameterization is to our knowledge the only existing constructive method to produce Bures-distributed states, and is straightforward to implement given its reliance on independent normal parameters only.

Following Bayes' rule, the posterior distribution becomes
\begin{widetext}
\begin{equation}
\label{eq:posterior}
\pi(\bx) = \frac{1}{\mathcal{Z}} L_\bbD(\bx) \pi_0 (\bx) \propto \left( \prod_{s=1}^{Rd^2} e^{-K(z) p_s(\by)} [K(z)p_s(\by)]^{N_s} \right)
 \left(\prod_{k=1}^{2d^4}e^{-\frac{1}{2}|y_k|^2} \right) e^{-\frac{1}{2}z^2},
\end{equation}
\end{widetext}
where $\mathcal{Z}$ is a constant such that $\int d\bx\,\pi(\bx) = 1$. We have adopted this notation for Bayes' theorem---rather than the more traditional $\Pr(\bx|\bbD)= \Pr(\bbD|\bx)\Pr(\bx)/\Pr(\bbD)$---to emphasize the functional dependencies on $\bx$, which are all that must be accounted for in the sampling algorithm below. From $\pi(\bx)$, the Bayesian mean estimator $f_B$ of any quantity (scalar, vector, or matrix) expressible as a function of $\bx$ can be estimated as 
\begin{equation}
\label{eq:BME}
f_B = \int d\bx\, \pi(\bx) f(\bx) \approx \frac{1}{S}\sum_{j=1}^S f(\bx^{(j)}),
\end{equation}
where, in lieu of direct integration, $S$ samples $\{\bx^{(1)},...,\bx^{(S)}\}$ are obtained from the distribution $\pi(\bx)$ through Markov chain Monte Carlo (MCMC) techniques, as described below.

\subsection{MCMC Sampling}
Acquiring the samples necessary for computation of high-dimensional integrals of the form in Eq.~\eqref{eq:BME} forms the primary bottleneck in Bayesian inference. The most common family of solutions to address this challenge fall under the general umbrella of MCMC, in which samples from a well chosen Markov chain are designed to approach the statistics of $\pi(\bx)$ asymptotically~\cite{Robert1999, MacKay2003}. Recently, we applied a particularly efficient MCMC algorithm---known as preconditioned Crank--Nicolson (pCN)~\cite{Cotter2013}---to the problem of Bayesian QST, finding significant computational improvements over previous implementations~\cite{Lukens2020b}. We utilize the same pCN approach here; the only difference from the algorithm in \cite{Lukens2020b} is a simpler acceptance probability depending exclusively on the likelihood ratio
, a consequence of the fact that all parameters here have normally distributed priors
~\cite{Cotter2013}.

\begin{figure*}[tb]
	\centering
	\includegraphics[trim=0 170 0 0,clip,width=7in]{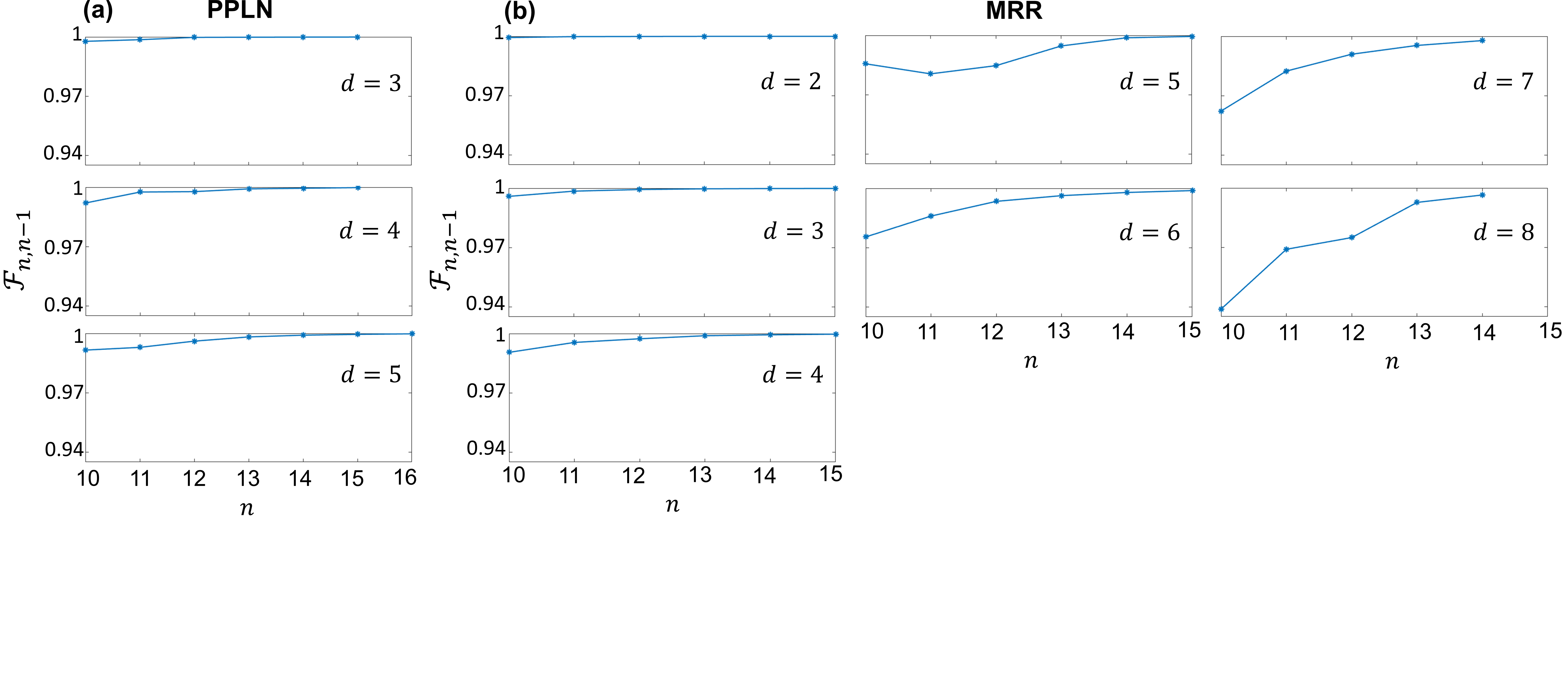}
	\caption{Sequential fidelity vs. thinning factor $n=\log_2 T$ for all dimensions for BFC states generated from (a) PPLN and (b) MRR.}
	\label{fig6}
\end{figure*}

For each collection of measurement results considered in the main text, we retain $S=2^{10}$ samples $\{(\by^{(j)},z^{(j)})\}$ from an MCMC chain of total length $ST$, where $T$ is a thinning factor that is successively doubled until convergence is obtained. From these samples, we then compute the Bayesian mean estimate of the density matrix
\begin{equation}
\label{eq:rhoBME}
\rho_B = \frac{1}{S} \sum_{j=1}^S \rho(\by^{(j)}),
\end{equation}
examples of which are plotted in Figs.~\ref{fig2}--\ref{fig4}. The mean and standard deviation of fidelity with respect to some ideal state $\ket{\Psi_d}$ can also be computed as
\begin{equation}
\label{eq:mean}
\overline{\cF_d} = \frac{1}{S} \sum_{j=1}^S \braket{\Psi_d|\rho(\by^{(j)})|\Psi_d}
\end{equation}
and
\begin{equation}
\label{eq:std}
\Delta\cF_d = \left( \frac{1}{S} \sum_{j=1}^S [\braket{\Psi_d|\rho(\by^{(j)})|\Psi_d}]^2 - \overline{\cF_d}^2 \right)^\frac{1}{2},
\end{equation}
respectively. For the PPLN states specifically, which are not dispersion-compensated, we take the state for comparison as
\begin{equation}
\label{eq:psiPPLN}
\ket{\Psi_d} = \frac{1}{\sqrt{d}} \sum_{m=1}^d e^{\imag\beta_2 L \Delta\omega^2m^2}\ket{m,m},
\end{equation}
with $\beta_2=2.06\times10^{-2}$ ps$^2$ m$^{-1}$ for standard single-mode fiber and $L=20$~m. The phases of the MRR states are precompensated, so for computing fidelities for them we use the uniform-phase version
\begin{equation}
\label{eq:psiMRR}
\ket{\Psi_d} = \frac{1}{\sqrt{d}} \sum_{m=1}^d \ket{m,m}.
\end{equation}


In order to monitor convergence with $T$, we consider the sequential fidelity, defined for a given $T=2^n$ as \begin{equation}
\label{eq:seq}
\cF_{n,n-1} = \left(\Tr \sqrt{\sqrt{\rho_B^{(n-1)}} \rho_B^{(n)} \sqrt{\rho_B^{(n-1)}}} \right)^2,
\end{equation}
where we use the notation $\rho_B^{(n)}$ to denote the Bayesian mean estimate [Eq.~\eqref{eq:BME}] for a chain of length $ST^n$.
For a sufficiently large thinning value $\cF_{n,n-1}$ will converge to unity and remain for all subsequent $n$. We note that this metric contains no reference to the ideal state $\ket{\Psi_d}$, but checks for consistency between subsequent Bayesian estimates only.

Figure~\ref{fig6} plots these sequential fidelities as a function of thinning ($n=\log_2 T$) for all BFCs characterized in the main text. In all cases, we consider the full measurement sets, $R=21$ for the PPLN results and $R=30$ for the MRR findings, as these cases generally possess the slowest MCMC convergence. 
As expected, the sequential fidelity converges more rapidly with $T$ for lower $d$; nonetheless, all examples ultimately reach $\cF_{n,n-1}>0.99$ at their respective maximum of $n$, indicating high MCMC convergence for all reported $\rho_B$ matrices. Continuing to even larger values of $T$ would certainly be desirable, particularly for $d=7$ and $d=8$, yet we are currently limited by computational power. For example, the $d=8$ MCMC chain with $T=2^{14}$ required almost a week to complete; with a total of 16,385 parameters to infer ($4d^4+1$), it is no surprise that we are pushing the limits of our desktop computer. Indeed, to our knowledge the present results with $d^2=64$ represent a record-high Hilbert-space dimension for complete Bayesian QST with a fully general (mixed state) prior, for any quantum system, simultaneously implying the efficiency of our existing MCMC algorithm as well as the importance of pursuing highly parallelizable MCMC methods~\cite{Jacob2020} to reach even larger dimensions in the future.

\subsection{Theoretical fidelities and ebits}
In our quantum system, several sources of noise, such as multipair emission and dark counts, are expected to be uniform across the BFC bins. 
Accordingly, as a simple model, we theoretically anticipate a ground truth quantum state of the form
\begin{equation}
\rho_\lambda = \lambda\ket{\Psi_d}\bra{\Psi_d} + \frac{1-\lambda}{d^2} I_{d^2},
\end{equation}
where $\ket{\Psi_d}$ is the ideal maximally entangled state [Eq.~\eqref{eq:psiPPLN} or \eqref{eq:psiMRR}],  $I_{d^2}$ is the $d^2 \times d^2$ identity operator, and $\lambda\in[0,1]$ determines the noise level. In this section, we provide quantities of interest (CAR, fidelity $\cF_d$, and log-negativity $E_d$) for $\rho_\lambda$ in order to compare against those found in Bayesian estimation: 
\begin{equation}
\label{eq:theory}
\begin{gathered}
\mathrm{CAR} = \frac{\max_{m} \braket{m,m|\rho_\lambda|m,m}}{\min_{m,n}\braket{m,n|\rho_\lambda|m,n}} = 1+\frac{d\lambda}{(1-\lambda)}  \\
\cF_d = \braket{\Psi_d | \rho_{\lambda} |\Psi_d} =\frac{(d^2-1)\lambda+1}{d^2} \\
E_d  = \begin{cases} 0 & ; \;\;  \cF_d < 1/d \\
\log_2 d \cF_d & ; \;\; \cF_d > 1/d
\end{cases},
\end{gathered}
\end{equation}
where the last formula is adapted from \cite{Vidal2002}.
The fidelities and log-negativities reported for $\rho_{\lambda}$ in the main text are calculated using these equations.
In general, both fidelity and $E_d$ deviate more strongly from their respective ideals as $d$ increases for a fixed CAR.
The good agreement in the main text between the Bayesian-estimated states and this simple white noise model suggests that our understanding of noise processes in the system is well justified.

\subsection{Theoretical analysis of measurement efficiency}

In this section we provide a basic theoretical analysis of the effectiveness of our proposed
tomographic method.
We first consider a highly simplified model that captures the key
aspects of how phase modulation and mode mixing enable tomographic
reconstruction, concentrating in this case on a single frequency-bin qudit occupying $d$ modes $x\in\{1,...,d\}$. While the experiments in the main text examine the joint state of \emph{two} qudits instead, this simpler single-qudit case reveals the basic principles of the tomography method with minimal distractions.  
The only type of measurement
considered is projection onto the photon's output mode, reflecting the condition of frequency-resolved detection; the probability
of obtaining bin $x$ is $p_{x}^{(0)}\equiv\rho_{xx}$, and the qudit assumption ensures that 
$\rho_{xx}=0$ for $x<1$ or $x>d$.

\begin{figure*}[tb!]
    \centering
    \includegraphics[width=5.5in]{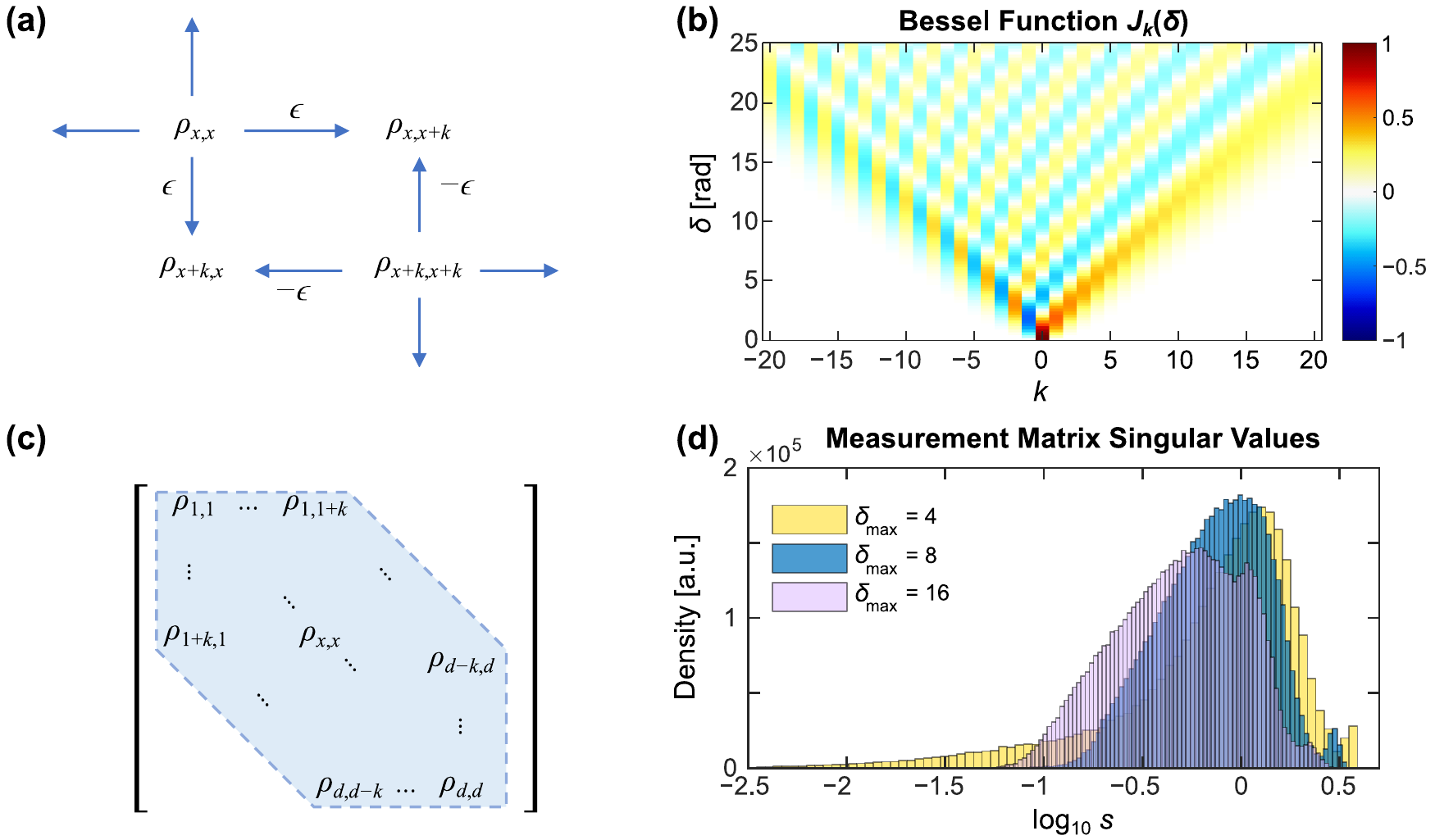}
    \caption{Theoretical analysis of measurement approach. (a)~Effect of the mode-mixing operator $S_k$ [Eq.~\eqref{eq:S_k}] on the density operator $\rho$. (b)~Behavior of the Bessel functions, which describe the modulation operators $T_\delta$ [Eq.~\eqref{eq: T_delta}]. (c)~The modulation operator $T_\delta$ with $\delta \approx k$ probes a $k$-band of $\rho$, i.e., the elements at most $k$ above or below the diagonal. (d)~Distribution of singular values for random measurement matrices $O$ for $d=8$,  $\delta \in [0,\delta_{\max}]$, and $R = 2d$: 
    (gold)  $\delta_{max}=4$; (blue) $\delta_{max}=8$; (violet) $\delta_{max}=16$.
    }
    \label{fig:combineFour}
\end{figure*}

Consider the mode mixing operator $S_{k}$ which weakly mixes each
mode $x$ with modes $x+k$ and $x-k$: 
\begin{align}
\label{eq:S_k}
S_{k}\ket x & =-\epsilon\ket{x-k}+\ket x+\epsilon\ket{x+k}.
\end{align}
For $k=1$, this approximates the action of an EOM with small modulation depth
$\delta=2\epsilon$ and modulation frequency equal to the fundamental mode spacing. More generally, for small $\epsilon$, $S_{k}$
is approximately unitary and describes a weak, translation-invariant
mode mixing operation [Fig.~\ref{fig:combineFour}(a)].

Suppose we apply $S_{k}$ to the photon and measure its bin. The
probability of observing bin $x$ is
\begin{align}
p_{x}^{(k)} & \equiv\bra xS_{k}\rho S_{k}^{\dagger}\ket x.
\end{align}
Using $S_{k}^{\dagger}\ket{x} =\epsilon\ket{x-k}+\ket x-\epsilon\ket{x+k}$ and the fact that $\rho_{yx}=\rho_{xy}^{*}$ we obtain
\begin{widetext}
\begin{equation}
\label{eq:real}
\begin{aligned}
p_{x}^{(k)} & =\left(\epsilon\bra{x-k}+\bra x-\epsilon\bra{x+k}\right)\rho\left(\epsilon\ket{x-k}+\ket x-\epsilon\ket{x+k}\right)\\
 & =\rho_{xx}+\epsilon\left(\rho_{x-k,x}+\rho_{x,x-k}-\rho_{x+k,x}-\rho_{x,x+k}\right)+\mathcal{O}(\epsilon^{2})\\
 & =\rho_{xx}+2\epsilon(\Re\rho_{x-k,x}-\Re\rho_{x,x+k})+\mathcal{O}(\epsilon^{2}).
\end{aligned}
\end{equation}

Now, from the fact that $|\rho_{xy}|^{2}\le\rho_{xx}\rho_{yy}$ and
the fact that the photon was initially restricted to modes $\{1,\ldots,d\}$
we have that the support of $\rho_{x,x+k}$ is $x\in\{1,...,d-k\}$ and
the support of $\rho_{x-k,x}$ is $x\in\{k+1,...,d\}$. Considering just
the outcomes in bins $\{1,\ldots,d\}$ yields the approximate system
of equations 
\begin{equation}
\begin{aligned}
-\Re\rho_{1,1+k} & =\frac{1}{2\epsilon}(p_{1}^{(k)}-p_{1}^{(0)})\\
 & \vdots\\
-\Re\rho_{k,2k} & =\frac{1}{2\epsilon}(p_{k}^{(k)}-p_{k}^{(0)})\\
\Re\rho_{1,k+1}-\Re\rho_{k+1,2k+1} & =\frac{1}{2\epsilon}(p_{k+1}^{(k)}-p_{k+1}^{(0)})\\
 & \vdots\\
\Re\rho_{d-2k,d-k}-\Re\rho_{d-k,d} & =\frac{1}{2\epsilon}(p_{d-k}^{(k)}-p_{d-k}^{(0)})\\
\Re\rho_{d-2k+1,d-k+1} & =\frac{1}{2\epsilon}(p_{d-k+1}^{(k)}-p_{d-k+1}^{(0)})\\
 & \vdots\\
\Re\rho_{d-k,d} & =\frac{1}{2\epsilon}(p_{d}^{(k)}-p_{d}^{(0)})
\end{aligned}
\end{equation}
valid for $k \in\{ 1,\ldots,d-1\}$ and small $\epsilon$. In principle, this (generally overdetermined) system of $d$ equations
enables one to determine the real parts of $\rho_{1,k+1},\ldots,\rho_{d-k,d}$,
the nonzero part of the $k$th diagonal of $\rho$. (Note that the $-k$th
diagonal of $\rho$ is just the conjugate of the $k$th diagonal.)
The elements on diagonals $\{-k,\ldots,0,\ldots,k\}$ will be collectively referred to as the $k$-band.

To determine the imaginary parts of the $k$th diagonal, each pair
of components $\rho_{x,x}$ and $\rho_{x+k,x}$ must be mixed with
a relative phase of $\pi/2$. Let $\Phi_k$ be the operation
\begin{align}
\Phi_{k}\ket x & =\varpi_{k}^{x}\ket x
\end{align}
where $\varpi_k=e^{\imag\pi/2k}$ is the $k$th principle root of $\imag$.
Suppose we apply $\Phi_{k}$ and then $S_{k}$ to the photon and measure
its mode. The probability of observing bin $x$ in this case is
\begin{equation}
\begin{aligned}
p_{x}^{(k)\prime} & \equiv\bra xS_{k}\Phi_{k}\rho\Phi_{k}^{\dagger}S_{k}^{\dagger}\ket x\\
 & =\bra xS_{k}\rho^{\prime}S_{k}^{\dagger}\ket x
\end{aligned}
\end{equation}
where $\rho^{\prime}=\Phi_{k}\rho\Phi_{k}^{\dagger}$. Using $\rho_{xy}^{\prime}=\varpi_{k}^{x-y}\rho_{xy}$
we have
\begin{equation}
\begin{aligned}
p_{x}^{(k)\prime} & =\left(\epsilon\ket{x-k}+\bra x-\epsilon\bra{x+k}\right)\rho^{\prime}\left(\epsilon\ket{x-k}+\ket x-\epsilon\ket{x+k}\right)\\
 & =\rho_{xx}^{\prime}+\epsilon\left(\rho_{x-k,x}^{\prime}+\rho_{x,x-k}^{\prime}-\rho_{x+k,k}-\rho_{x,x+k}\right)+\mathcal{O}(\epsilon^{2})\\
 & =\rho_{xx}+\epsilon\left(\varpi_{k}^{-k}\rho_{x-k,x}+\varpi_{k}^{k}\rho_{x,x-k}-\varpi_{k}^{k}\rho_{x+k,k}-\varpi_{k}^{-k}\rho_{x,x+k}\right)+\mathcal{O}(\epsilon^{2}).
\end{aligned}
\end{equation}
Since $\varpi_{k}^{\pm k}=\pm\imag$ we have
\begin{equation}
\label{eq:imag}
\begin{aligned}
p_{x}^{(k)\prime} & =\rho_{xx}^{\prime}+\epsilon\left(-\imag\rho_{x-k,x}+\imag\rho_{x,x-k}-\imag\rho_{x+k,k}+\imag\rho_{x,x+k}\right)+\mathcal{O}(\epsilon^{2})\\
 & =\rho_{xx}+2\epsilon(\Im\rho_{x-k,x}-\Im\rho_{x,x+k})+\mathcal{O}(\epsilon^{2}).
\end{aligned}
\end{equation}
\end{widetext}
Using the probabilities of bins 1 through $d$ we obtain a system
of $d$ equations which, together with $p^{(0)}$, in principle enables
determination of the imaginary parts of the $k$th diagonal.

To summarize: $p^{(0)}$ determines the diagonal elements of $\rho$,
then $p^{(k)}$ and $p^{(k)\prime}$ (obtained by applying $S_{k}$
and $S_{k}\Phi_{k}$, respectively) determine the off-diagonal elements
of $\rho$ that lie $k$ positions above and below the diagonal. Thus,
all $d^2$ elements of the single-qudit density matrix can be probed by $1 + 2(d-1) = 2d-1$ experimental settings with $d$ outcomes each. 

To relate this to the experimental approach, we note that the EOM
effects a mode mixing operation $T_{\delta}$ of the form [cf. Eq.~\eqref{eq:unit2}]
\begin{align}
\label{eq: T_delta}
T_{\delta}\ket x & =\sum_{k=-\infty}^{\infty}J_{k}(\delta)\ket{x+k}
\end{align}
As shown in Fig.~\ref{fig:combineFour}(b), $J_{k}(\delta)$ is oscillatory
in $k$ with approximate support $k\in\{-\lceil\delta\rceil,\ldots,\lceil\delta\rceil\}$. That
is, $T_{\delta}$ is approximately a linear combination of $\{S_{1},\ldots,S_{\lceil\delta\rceil}\}$,
mixing each $x$ with multiple modes ranging from roughly $x-\lceil\delta\rceil$ to
$x+\lceil\delta\rceil$ and yielding information about the elements in the $\lceil\delta\rceil$-band
of $\rho$ [Fig.~\ref{fig:combineFour}(c)]. If the elements in the $(\lceil\delta\rceil-1)$-band have already
been determined, $T_{\delta}$ provides new information
primarily concerning the $\lceil\delta\rceil$-band.
In our specific experiments, we applied a set of random phases with the pulse shaper in addition to the modulation, allowing us to probe linear combinations of both the real and imaginary parts of the $k$-band similar to the $S_k$ and $\Phi_{k}$ operators in the discussion above.
In this way, a collection of measurements with modulation indices uniformly distributed $\delta\in[0, k]$ along with random phases is sufficient to fully probe the $k$-band. And since the complete $d\times d$ density matrix $\rho$ is encompassed within the ($d-1$)-band, this model suggests $\delta_{\max} \sim \mathcal{O}(d)$ as an ideal design choice for efficient state tomography with our method.

This intuitive picture reveals how an EOM-based mode mixer can be designed to respond to all elements of the density matrix, yet it does not quantify how well such measurements span the space of hermitian operators; if the mixing weights are small or produce excessive scattering into modes outside of the computational space, the number of observations required to reach a desired accuracy will be high, even if the measurements are tomographically complete. To address this question explicitly, we next explore the specifics of the measurement operations through singular value decomposition. 
Suppose we have a set of $R$ different measurements $\mathbf{M}_{1},\ldots,\mathbf{M}_{R}$,
each of which is described by a set of positive operator-valued measures
(POVM): $\mathbf{M}_{i}=\{M_{i1},\ldots M_{im_{i}}\}$ where $m_{i}$
is the number of possible outcomes of $\mathbf{M}_{i}$ ($d$ in our case) and each $M_{ij}$
is a positive semidefinite operator. To account for photon scattering outside of the measured computational space, the operators need not conserve probability.
When $\mathbf{M}_{i}$ is performed on state $\rho$, the probability
of outcome $j$ is
\begin{align}
p_{ij} & =\Tr(M_{ij}\rho).
\end{align}
This may be summarized as the linear system
\begin{align}
p & =O^{\dagger}\vec{\rho}\label{eq: p =00003D O rho}
\end{align}
where $\vec{\rho}$ is the vectorization of $\rho$ and 
\begin{align}
O & =\left[\begin{array}{ccccccc}
\vec{M}_{11} & \cdots & \vec{M}_{1m_{1}} & \cdots & \vec{M}_{R1} & \cdots & \vec{M}_{Rm_{R}}\end{array}\right].
\end{align}
is a $d^{2}\times m$ matrix where $m=m_{1}+\cdots+m_{R}$. The measurement
set $\mathcal{M}=\{\mathbf{M}_{1},\ldots,\mathbf{M}_{R}\}$ is called
\emph{informationally complete} if $\rho$ is uniquely determined
by measured probabilities $p$. This occurs iff $O$ is full rank
(i.e., rank $d^{2}$), which requires at least $d$ different measurements
($R\ge d$) with $d$ outcomes each. Practically speaking, however, more important than the attainment of informational completeness is the actual distribution of 
singular values $s_{1}\ge\cdots\ge s_{d^{2}}$ of $O$. If the singular values
are all of order 1, then $\mathcal{M}$ determines all components
of $\rho$ with comparable sensitivity. But any singular values of
$O$ that are much smaller than 1 correspond to components of $\rho$
to which $\mathcal{M}$ is only weakly sensitive. The estimates of such components will
be susceptible to statistical fluctuations in the experimental
data.

As examples, we compute singular values for a variety of settings at $d=8$---the maximum single-qudit dimension characterized experimentally. We consider $R=2d=16$ settings, which we found sufficient to obtain robust and repeatable distributions. 
Each measurement configuration involves a modulation index $\delta$ chosen uniformly at random in the interval $[0,\delta_{\max}$], preceded by a random phase vector $\vec{\phi}\in[0,2\pi]^{d}$ applied by the pulse shaper. Figure~\ref{fig:combineFour}(d) plots histograms of singular values from 2,000 measurement matrices $O$ each for $\delta_{\max}\in\{4,8,16\}$, corresponding to $d/2$, $d$, and $2d$ at $d=8$. A narrow peak around $\log_{10} s \approx 0$ indicates high and comparable sensitivity for all elements in the density matrix; indeed, a complete set of mutually unbiased bases---an ideal choice for tomography~\cite{Wootters1989}---possesses $d^2$ equal singular values at $s=1$. By this criterion, $\delta_{\max}=d=8$ is seen to provide the most efficient measurement distribution of the three examples. For the smaller index of $\delta_{\max}=4$, the main peak is accompanied by a strong tail indicating small sensitivity to an appreciable percentage of the matrix elements. And for the larger index $\delta_{\max}=16$, the main peak shifts to lower values, which reflects ``over-modulating'' of the quantum state; taking $\delta_{\max}$ beyond $d$ increases the probability of scattering the input outside of the computational space [see Fig.~\ref{fig:combineFour}(c)] without any improvement to mixing within the space. Our numerical findings therefore join the simple theory above in suggesting a maximum modulation index of $\delta_{\max}\approx d$. Nevertheless, we note that all cases in Fig.~\ref{fig:combineFour}(d) sample the Hilbert space comprehensively, so that any $\delta_{\max}\sim\mathcal{O}(d)$ is likely to prove sufficient in a tomographic context. Ultimately, we emphasize that the theory developed here is meant to provide heuristic guidelines for implementing EOM-based frequency-bin tomography, not to imply the optimality of our specific measurements. For example, it is quite possible that alternative distributions for $\delta$---e.g., other than the uniform draw $\delta\in[0,\delta_{\max}]$---may show more favorable properties with further research. Nevertheless, our reliance on Bayesian estimation ensures that these questions need not be answered for useful inference.

Extending to the experimental task of two-qudit characterization, our model thus indicates one should consider joint measurements of the form $T_{\delta_1} \otimes T_{\delta_2}$ with $\delta_{1}, \delta_2 \in [0, d]$. 
Considering our experimental values of $\delta_{\max}=2.5$ for the PPLN tests and 3.4 for the MRR, $\delta_{\max}$ falls in the range of $0.425d$ and $1.7d$ for all qudit dimensions examined, aligning well with the $\mathcal{O}(d)$ desideratum. Since the entire density matrix of a $d^2$-dimensional two-qudit state requires specification of $d^4-1$ real parameters and each experimental pulse shaper/EOM setting $r$ provides $d^2$ outcomes, 
$R\sim\mathcal{O}(d^2)$ would be expected to be required to fully probe the two-qudit Hilbert space. Empirically, however, we were able to attain low-error state reconstruction with fewer measurements: e.g., $R=10$ instead of 25 for the $d=5$ PPLN BFC [Fig.~\ref{fig2}(a)] and $R=30$ instead of 64 for the $d=8$ MRR BFC [Fig.~\ref{fig4}(b)]. Several aspects are likely responsible for this reduction. From a broad perspective, the positive semidefiniteness of the density matrix imposes additional constraints that are not reflected in a linear system analysis but automatically accounted for in the Bayesian inference procedure. Thus, simply comparing the number of measurements with the number of parameters generally yields an overly pessimistic assessment of the information required for tomography.

Moreover, we suspect that specific features of our quantum state also contribute to more efficient reconstruction. The highly correlated nature revealed in the first JSI measurement implies that only off-diagonal elements of the form $\rho_{(xx)(x+k,x+k)}$---i.e., those satisfying biphoton energy conservation---can be significantly different than zero. This reduces the number of appreciable elements in our density matrix from $\mathcal{O}(d^4)$ to $\mathcal{O}(d^2)$. 
Of course, the other off-diagonal density matrix elements are not strictly zero in practice, and the Bayesian model neither requires nor assumes such a simplification. However, the strong frequency correlations do eliminate a large portion of the Hilbert space from consideration. Consequently, future experiments with more general states 
may require more experimental settings for low-uncertainty estimation than we have currently used.

\textbf{Data Availability}

The datasets generated during and/or analysed during the current study are available from the corresponding author on reasonable request.
\smallskip

\textbf{Code Availability}

Analysis code used in this study is available from the corresponding author on request.


\begin{thebibliography}{46}%
\makeatletter
\providecommand \@ifxundefined [1]{%
 \@ifx{#1\undefined}
}%
\providecommand \@ifnum [1]{%
 \ifnum #1\expandafter \@firstoftwo
 \else \expandafter \@secondoftwo
 \fi
}%
\providecommand \@ifx [1]{%
 \ifx #1\expandafter \@firstoftwo
 \else \expandafter \@secondoftwo
 \fi
}%
\providecommand \natexlab [1]{#1}%
\providecommand \enquote  [1]{``#1''}%
\providecommand \bibnamefont  [1]{#1}%
\providecommand \bibfnamefont [1]{#1}%
\providecommand \citenamefont [1]{#1}%
\providecommand \href@noop [0]{\@secondoftwo}%
\providecommand \href [0]{\begingroup \@sanitize@url \@href}%
\providecommand \@href[1]{\@@startlink{#1}\@@href}%
\providecommand \@@href[1]{\endgroup#1\@@endlink}%
\providecommand \@sanitize@url [0]{\catcode `\\12\catcode `\$12\catcode
  `\&12\catcode `\#12\catcode `\^12\catcode `\_12\catcode `\%12\relax}%
\providecommand \@@startlink[1]{}%
\providecommand \@@endlink[0]{}%
\providecommand \url  [0]{\begingroup\@sanitize@url \@url }%
\providecommand \@url [1]{\endgroup\@href {#1}{\urlprefix }}%
\providecommand \urlprefix  [0]{URL }%
\providecommand \Eprint [0]{\href }%
\providecommand \doibase [0]{http://dx.doi.org/}%
\providecommand \selectlanguage [0]{\@gobble}%
\providecommand \bibinfo  [0]{\@secondoftwo}%
\providecommand \bibfield  [0]{\@secondoftwo}%
\providecommand \translation [1]{[#1]}%
\providecommand \BibitemOpen [0]{}%
\providecommand \bibitemStop [0]{}%
\providecommand \bibitemNoStop [0]{.\EOS\space}%
\providecommand \EOS [0]{\spacefactor3000\relax}%
\providecommand \BibitemShut  [1]{\csname bibitem#1\endcsname}%
\let\auto@bib@innerbib\@empty
\bibitem [{\citenamefont {Kues}\ \emph {et~al.}(2019)\citenamefont {Kues},
  \citenamefont {Reimer}, \citenamefont {Lukens}, \citenamefont {Munro},
  \citenamefont {Weiner}, \citenamefont {Moss},\ and\ \citenamefont
  {Morandotti}}]{Kues2019}%
  \BibitemOpen
  \bibfield  {author} {\bibinfo {author} {\bibfnamefont {M.}~\bibnamefont
  {Kues}}, \bibinfo {author} {\bibfnamefont {C.}~\bibnamefont {Reimer}},
  \bibinfo {author} {\bibfnamefont {J.~M.}\ \bibnamefont {Lukens}}, \bibinfo
  {author} {\bibfnamefont {W.~J.}\ \bibnamefont {Munro}}, \bibinfo {author}
  {\bibfnamefont {A.~M.}\ \bibnamefont {Weiner}}, \bibinfo {author}
  {\bibfnamefont {D.~J.}\ \bibnamefont {Moss}}, \ and\ \bibinfo {author}
  {\bibfnamefont {R.}~\bibnamefont {Morandotti}},\ }\href@noop {} {\bibfield
  {journal} {\bibinfo  {journal} {Nat. Photonics}\ }\textbf {\bibinfo {volume}
  {13}},\ \bibinfo {pages} {170} (\bibinfo {year} {2019})}\BibitemShut
  {NoStop}%
\bibitem [{\citenamefont {Kues}\ \emph {et~al.}(2017)\citenamefont {Kues},
  \citenamefont {Reimer}, \citenamefont {Roztocki}, \citenamefont {Cort\'{e}s},
  \citenamefont {Sciara}, \citenamefont {Wetzel}, \citenamefont {Zhang},
  \citenamefont {Cino}, \citenamefont {Chu}, \citenamefont {Little},
  \citenamefont {Moss}, \citenamefont {Caspani}, \citenamefont {Aza\~{n}a},\
  and\ \citenamefont {Morandotti}}]{Kues2017}%
  \BibitemOpen
  \bibfield  {author} {\bibinfo {author} {\bibfnamefont {M.}~\bibnamefont
  {Kues}}, \bibinfo {author} {\bibfnamefont {C.}~\bibnamefont {Reimer}},
  \bibinfo {author} {\bibfnamefont {P.}~\bibnamefont {Roztocki}}, \bibinfo
  {author} {\bibfnamefont {L.~R.}\ \bibnamefont {Cort\'{e}s}}, \bibinfo
  {author} {\bibfnamefont {S.}~\bibnamefont {Sciara}}, \bibinfo {author}
  {\bibfnamefont {B.}~\bibnamefont {Wetzel}}, \bibinfo {author} {\bibfnamefont
  {Y.}~\bibnamefont {Zhang}}, \bibinfo {author} {\bibfnamefont
  {A.}~\bibnamefont {Cino}}, \bibinfo {author} {\bibfnamefont {S.~T.}\
  \bibnamefont {Chu}}, \bibinfo {author} {\bibfnamefont {B.~E.}\ \bibnamefont
  {Little}}, \bibinfo {author} {\bibfnamefont {D.~J.}\ \bibnamefont {Moss}},
  \bibinfo {author} {\bibfnamefont {L.}~\bibnamefont {Caspani}}, \bibinfo
  {author} {\bibfnamefont {J.}~\bibnamefont {Aza\~{n}a}}, \ and\ \bibinfo
  {author} {\bibfnamefont {R.}~\bibnamefont {Morandotti}},\ }\href@noop {}
  {\bibfield  {journal} {\bibinfo  {journal} {Nature}\ }\textbf {\bibinfo
  {volume} {546}},\ \bibinfo {pages} {622} (\bibinfo {year}
  {2017})}\BibitemShut {NoStop}%
\bibitem [{\citenamefont {Imany}\ \emph {et~al.}(2018)\citenamefont {Imany},
  \citenamefont {Jaramillo-Villegas}, \citenamefont {Odele}, \citenamefont
  {Han}, \citenamefont {Leaird}, \citenamefont {Lukens}, \citenamefont
  {Lougovski}, \citenamefont {Qi},\ and\ \citenamefont {Weiner}}]{Imany2018}%
  \BibitemOpen
  \bibfield  {author} {\bibinfo {author} {\bibfnamefont {P.}~\bibnamefont
  {Imany}}, \bibinfo {author} {\bibfnamefont {J.~A.}\ \bibnamefont
  {Jaramillo-Villegas}}, \bibinfo {author} {\bibfnamefont {O.~D.}\ \bibnamefont
  {Odele}}, \bibinfo {author} {\bibfnamefont {K.}~\bibnamefont {Han}}, \bibinfo
  {author} {\bibfnamefont {D.~E.}\ \bibnamefont {Leaird}}, \bibinfo {author}
  {\bibfnamefont {J.~M.}\ \bibnamefont {Lukens}}, \bibinfo {author}
  {\bibfnamefont {P.}~\bibnamefont {Lougovski}}, \bibinfo {author}
  {\bibfnamefont {M.}~\bibnamefont {Qi}}, \ and\ \bibinfo {author}
  {\bibfnamefont {A.~M.}\ \bibnamefont {Weiner}},\ }\href@noop {} {\bibfield
  {journal} {\bibinfo  {journal} {Opt. Express}\ }\textbf {\bibinfo {volume}
  {26}},\ \bibinfo {pages} {1825} (\bibinfo {year} {2018})}\BibitemShut
  {NoStop}%
\bibitem [{\citenamefont {Lu}\ \emph {et~al.}(2018{\natexlab{a}})\citenamefont
  {Lu}, \citenamefont {Lukens}, \citenamefont {Peters}, \citenamefont
  {Williams}, \citenamefont {Weiner},\ and\ \citenamefont
  {Lougovski}}]{Lu2018b}%
  \BibitemOpen
  \bibfield  {author} {\bibinfo {author} {\bibfnamefont {H.-H.}\ \bibnamefont
  {Lu}}, \bibinfo {author} {\bibfnamefont {J.~M.}\ \bibnamefont {Lukens}},
  \bibinfo {author} {\bibfnamefont {N.~A.}\ \bibnamefont {Peters}}, \bibinfo
  {author} {\bibfnamefont {B.~P.}\ \bibnamefont {Williams}}, \bibinfo {author}
  {\bibfnamefont {A.~M.}\ \bibnamefont {Weiner}}, \ and\ \bibinfo {author}
  {\bibfnamefont {P.}~\bibnamefont {Lougovski}},\ }\href@noop {} {\bibfield
  {journal} {\bibinfo  {journal} {Optica}\ }\textbf {\bibinfo {volume} {5}},\
  \bibinfo {pages} {1455} (\bibinfo {year} {2018}{\natexlab{a}})}\BibitemShut
  {NoStop}%
\bibitem [{\citenamefont {Blume-Kohout}(2010)}]{Blume2010}%
  \BibitemOpen
  \bibfield  {author} {\bibinfo {author} {\bibfnamefont {R.}~\bibnamefont
  {Blume-Kohout}},\ }\href {http://stacks.iop.org/1367-2630/12/i=4/a=043034}
  {\bibfield  {journal} {\bibinfo  {journal} {New J. Phys.}\ }\textbf {\bibinfo
  {volume} {12}},\ \bibinfo {pages} {043034} (\bibinfo {year}
  {2010})}\BibitemShut {NoStop}%
\bibitem [{\citenamefont {Lukens}\ \emph
  {et~al.}(2020{\natexlab{a}})\citenamefont {Lukens}, \citenamefont {Law},
  \citenamefont {Jasra},\ and\ \citenamefont {Lougovski}}]{Lukens2020b}%
  \BibitemOpen
  \bibfield  {author} {\bibinfo {author} {\bibfnamefont {J.~M.}\ \bibnamefont
  {Lukens}}, \bibinfo {author} {\bibfnamefont {K.~J.~H.}\ \bibnamefont {Law}},
  \bibinfo {author} {\bibfnamefont {A.}~\bibnamefont {Jasra}}, \ and\ \bibinfo
  {author} {\bibfnamefont {P.}~\bibnamefont {Lougovski}},\ }\href@noop {}
  {\bibfield  {journal} {\bibinfo  {journal} {New J. Phys.}\ }\textbf {\bibinfo
  {volume} {22}},\ \bibinfo {pages} {063038} (\bibinfo {year}
  {2020}{\natexlab{a}})}\BibitemShut {NoStop}%
\bibitem [{\citenamefont {Erhard}\ \emph {et~al.}(2020)\citenamefont {Erhard},
  \citenamefont {Krenn},\ and\ \citenamefont {Zeilinger}}]{Erhard2020}%
  \BibitemOpen
  \bibfield  {author} {\bibinfo {author} {\bibfnamefont {M.}~\bibnamefont
  {Erhard}}, \bibinfo {author} {\bibfnamefont {M.}~\bibnamefont {Krenn}}, \
  and\ \bibinfo {author} {\bibfnamefont {A.}~\bibnamefont {Zeilinger}},\
  }\href@noop {} {\bibfield  {journal} {\bibinfo  {journal} {Nat. Rev. Phys.}\
  }\textbf {\bibinfo {volume} {2}},\ \bibinfo {pages} {365} (\bibinfo {year}
  {2020})}\BibitemShut {NoStop}%
\bibitem [{\citenamefont {Cozzolino}\ \emph {et~al.}(2019)\citenamefont
  {Cozzolino}, \citenamefont {Da~Lio}, \citenamefont {Bacco},\ and\
  \citenamefont {Oxenl{\o}we}}]{Cozzolino2019}%
  \BibitemOpen
  \bibfield  {author} {\bibinfo {author} {\bibfnamefont {D.}~\bibnamefont
  {Cozzolino}}, \bibinfo {author} {\bibfnamefont {B.}~\bibnamefont {Da~Lio}},
  \bibinfo {author} {\bibfnamefont {D.}~\bibnamefont {Bacco}}, \ and\ \bibinfo
  {author} {\bibfnamefont {L.~K.}\ \bibnamefont {Oxenl{\o}we}},\ }\href@noop {}
  {\bibfield  {journal} {\bibinfo  {journal} {Adv. Quantum Technol.}\ }\textbf
  {\bibinfo {volume} {2}},\ \bibinfo {pages} {1900038} (\bibinfo {year}
  {2019})}\BibitemShut {NoStop}%
\bibitem [{\citenamefont {Barreiro}\ \emph {et~al.}(2008)\citenamefont
  {Barreiro}, \citenamefont {Wei},\ and\ \citenamefont {Kwiat}}]{Barreiro2008}%
  \BibitemOpen
  \bibfield  {author} {\bibinfo {author} {\bibfnamefont {J.~T.}\ \bibnamefont
  {Barreiro}}, \bibinfo {author} {\bibfnamefont {T.-C.}\ \bibnamefont {Wei}}, \
  and\ \bibinfo {author} {\bibfnamefont {P.~G.}\ \bibnamefont {Kwiat}},\
  }\href@noop {} {\bibfield  {journal} {\bibinfo  {journal} {Nat. Phys.}\
  }\textbf {\bibinfo {volume} {4}},\ \bibinfo {pages} {282} (\bibinfo {year}
  {2008})}\BibitemShut {NoStop}%
\bibitem [{\citenamefont {Cerf}\ \emph {et~al.}(2002)\citenamefont {Cerf},
  \citenamefont {Bourennane}, \citenamefont {Karlsson},\ and\ \citenamefont
  {Gisin}}]{Cerf2002}%
  \BibitemOpen
  \bibfield  {author} {\bibinfo {author} {\bibfnamefont {N.~J.}\ \bibnamefont
  {Cerf}}, \bibinfo {author} {\bibfnamefont {M.}~\bibnamefont {Bourennane}},
  \bibinfo {author} {\bibfnamefont {A.}~\bibnamefont {Karlsson}}, \ and\
  \bibinfo {author} {\bibfnamefont {N.}~\bibnamefont {Gisin}},\ }\href@noop {}
  {\bibfield  {journal} {\bibinfo  {journal} {Phys. Rev. Lett.}\ }\textbf
  {\bibinfo {volume} {88}},\ \bibinfo {pages} {127902} (\bibinfo {year}
  {2002})}\BibitemShut {NoStop}%
\bibitem [{\citenamefont {Ecker}\ \emph {et~al.}(2019)\citenamefont {Ecker},
  \citenamefont {Bouchard}, \citenamefont {Bulla}, \citenamefont {Brandt},
  \citenamefont {Kohout}, \citenamefont {Steinlechner}, \citenamefont
  {Fickler}, \citenamefont {Malik}, \citenamefont {Guryanova}, \citenamefont
  {Ursin},\ and\ \citenamefont {Huber}}]{Ecker2019}%
  \BibitemOpen
  \bibfield  {author} {\bibinfo {author} {\bibfnamefont {S.}~\bibnamefont
  {Ecker}}, \bibinfo {author} {\bibfnamefont {F.}~\bibnamefont {Bouchard}},
  \bibinfo {author} {\bibfnamefont {L.}~\bibnamefont {Bulla}}, \bibinfo
  {author} {\bibfnamefont {F.}~\bibnamefont {Brandt}}, \bibinfo {author}
  {\bibfnamefont {O.}~\bibnamefont {Kohout}}, \bibinfo {author} {\bibfnamefont
  {F.}~\bibnamefont {Steinlechner}}, \bibinfo {author} {\bibfnamefont
  {R.}~\bibnamefont {Fickler}}, \bibinfo {author} {\bibfnamefont
  {M.}~\bibnamefont {Malik}}, \bibinfo {author} {\bibfnamefont
  {Y.}~\bibnamefont {Guryanova}}, \bibinfo {author} {\bibfnamefont
  {R.}~\bibnamefont {Ursin}}, \ and\ \bibinfo {author} {\bibfnamefont
  {M.}~\bibnamefont {Huber}},\ }\href {\doibase 10.1103/PhysRevX.9.041042}
  {\bibfield  {journal} {\bibinfo  {journal} {Phys. Rev. X}\ }\textbf {\bibinfo
  {volume} {9}},\ \bibinfo {pages} {041042} (\bibinfo {year}
  {2019})}\BibitemShut {NoStop}%
\bibitem [{\citenamefont {V{\'e}rtesi}\ \emph {et~al.}(2010)\citenamefont
  {V{\'e}rtesi}, \citenamefont {Pironio},\ and\ \citenamefont
  {Brunner}}]{vertesi2010closing}%
  \BibitemOpen
  \bibfield  {author} {\bibinfo {author} {\bibfnamefont {T.}~\bibnamefont
  {V{\'e}rtesi}}, \bibinfo {author} {\bibfnamefont {S.}~\bibnamefont
  {Pironio}}, \ and\ \bibinfo {author} {\bibfnamefont {N.}~\bibnamefont
  {Brunner}},\ }\href@noop {} {\bibfield  {journal} {\bibinfo  {journal} {Phys.
  Rev. Lett.}\ }\textbf {\bibinfo {volume} {104}},\ \bibinfo {pages} {060401}
  (\bibinfo {year} {2010})}\BibitemShut {NoStop}%
\bibitem [{\citenamefont {Wang}\ \emph {et~al.}(2018)\citenamefont {Wang},
  \citenamefont {Paesani}, \citenamefont {Ding}, \citenamefont {Santagati},
  \citenamefont {Skrzypczyk}, \citenamefont {Salavrakos}, \citenamefont {Tura},
  \citenamefont {Augusiak}, \citenamefont {Man{\v{c}}inska}, \citenamefont
  {Bacco}, \citenamefont {Bonneau}, \citenamefont {Silverstone}, \citenamefont
  {Gong}, \citenamefont {Ac{\'{\i}}n}, \citenamefont {Rottwitt}, \citenamefont
  {Oxenl{\o}we}, \citenamefont {O'Brien}, \citenamefont {Laing},\ and\
  \citenamefont {Thompson}}]{Wang2018}%
  \BibitemOpen
  \bibfield  {author} {\bibinfo {author} {\bibfnamefont {J.}~\bibnamefont
  {Wang}}, \bibinfo {author} {\bibfnamefont {S.}~\bibnamefont {Paesani}},
  \bibinfo {author} {\bibfnamefont {Y.}~\bibnamefont {Ding}}, \bibinfo {author}
  {\bibfnamefont {R.}~\bibnamefont {Santagati}}, \bibinfo {author}
  {\bibfnamefont {P.}~\bibnamefont {Skrzypczyk}}, \bibinfo {author}
  {\bibfnamefont {A.}~\bibnamefont {Salavrakos}}, \bibinfo {author}
  {\bibfnamefont {J.}~\bibnamefont {Tura}}, \bibinfo {author} {\bibfnamefont
  {R.}~\bibnamefont {Augusiak}}, \bibinfo {author} {\bibfnamefont
  {L.}~\bibnamefont {Man{\v{c}}inska}}, \bibinfo {author} {\bibfnamefont
  {D.}~\bibnamefont {Bacco}}, \bibinfo {author} {\bibfnamefont
  {D.}~\bibnamefont {Bonneau}}, \bibinfo {author} {\bibfnamefont {J.~W.}\
  \bibnamefont {Silverstone}}, \bibinfo {author} {\bibfnamefont
  {Q.}~\bibnamefont {Gong}}, \bibinfo {author} {\bibfnamefont {A.}~\bibnamefont
  {Ac{\'{\i}}n}}, \bibinfo {author} {\bibfnamefont {K.}~\bibnamefont
  {Rottwitt}}, \bibinfo {author} {\bibfnamefont {L.~K.}\ \bibnamefont
  {Oxenl{\o}we}}, \bibinfo {author} {\bibfnamefont {J.~L.}\ \bibnamefont
  {O'Brien}}, \bibinfo {author} {\bibfnamefont {A.}~\bibnamefont {Laing}}, \
  and\ \bibinfo {author} {\bibfnamefont {M.~G.}\ \bibnamefont {Thompson}},\
  }\href {\doibase 10.1126/science.aar7053} {\bibfield  {journal} {\bibinfo
  {journal} {Science}\ }\textbf {\bibinfo {volume} {360}},\ \bibinfo {pages}
  {285} (\bibinfo {year} {2018})}\BibitemShut {NoStop}%
\bibitem [{\citenamefont {Qiang}\ \emph {et~al.}(2018)\citenamefont {Qiang},
  \citenamefont {Zhou}, \citenamefont {Wang}, \citenamefont {Wilkes},
  \citenamefont {Loke}, \citenamefont {O'Gara}, \citenamefont {Kling},
  \citenamefont {Marshall}, \citenamefont {Santagati}, \citenamefont {Ralph},
  \citenamefont {Wang}, \citenamefont {O'Brien}, \citenamefont {Thompson},\
  and\ \citenamefont {Matthews}}]{Qiang2018}%
  \BibitemOpen
  \bibfield  {author} {\bibinfo {author} {\bibfnamefont {X.}~\bibnamefont
  {Qiang}}, \bibinfo {author} {\bibfnamefont {X.}~\bibnamefont {Zhou}},
  \bibinfo {author} {\bibfnamefont {J.}~\bibnamefont {Wang}}, \bibinfo {author}
  {\bibfnamefont {C.~M.}\ \bibnamefont {Wilkes}}, \bibinfo {author}
  {\bibfnamefont {T.}~\bibnamefont {Loke}}, \bibinfo {author} {\bibfnamefont
  {S.}~\bibnamefont {O'Gara}}, \bibinfo {author} {\bibfnamefont
  {L.}~\bibnamefont {Kling}}, \bibinfo {author} {\bibfnamefont {G.~D.}\
  \bibnamefont {Marshall}}, \bibinfo {author} {\bibfnamefont {R.}~\bibnamefont
  {Santagati}}, \bibinfo {author} {\bibfnamefont {T.~C.}\ \bibnamefont
  {Ralph}}, \bibinfo {author} {\bibfnamefont {J.~B.}\ \bibnamefont {Wang}},
  \bibinfo {author} {\bibfnamefont {J.~L.}\ \bibnamefont {O'Brien}}, \bibinfo
  {author} {\bibfnamefont {M.~G.}\ \bibnamefont {Thompson}}, \ and\ \bibinfo
  {author} {\bibfnamefont {J.~C.~F.}\ \bibnamefont {Matthews}},\ }\href
  {\doibase 10.1038/s41566-018-0236-y} {\bibfield  {journal} {\bibinfo
  {journal} {Nat. Photonics}\ }\textbf {\bibinfo {volume} {12}},\ \bibinfo
  {pages} {534} (\bibinfo {year} {2018})}\BibitemShut {NoStop}%
\bibitem [{\citenamefont {Bavaresco}\ \emph {et~al.}(2018)\citenamefont
  {Bavaresco}, \citenamefont {Valencia}, \citenamefont {Kl{\"o}ckl},
  \citenamefont {Pivoluska}, \citenamefont {Erker}, \citenamefont {Friis},
  \citenamefont {Malik},\ and\ \citenamefont {Huber}}]{Bavaresco2018}%
  \BibitemOpen
  \bibfield  {author} {\bibinfo {author} {\bibfnamefont {J.}~\bibnamefont
  {Bavaresco}}, \bibinfo {author} {\bibfnamefont {N.~H.}\ \bibnamefont
  {Valencia}}, \bibinfo {author} {\bibfnamefont {C.}~\bibnamefont
  {Kl{\"o}ckl}}, \bibinfo {author} {\bibfnamefont {M.}~\bibnamefont
  {Pivoluska}}, \bibinfo {author} {\bibfnamefont {P.}~\bibnamefont {Erker}},
  \bibinfo {author} {\bibfnamefont {N.}~\bibnamefont {Friis}}, \bibinfo
  {author} {\bibfnamefont {M.}~\bibnamefont {Malik}}, \ and\ \bibinfo {author}
  {\bibfnamefont {M.}~\bibnamefont {Huber}},\ }\href@noop {} {\bibfield
  {journal} {\bibinfo  {journal} {Nat. Phys.}\ }\textbf {\bibinfo {volume}
  {14}},\ \bibinfo {pages} {1032} (\bibinfo {year} {2018})}\BibitemShut
  {NoStop}%
\bibitem [{\citenamefont {Brandt}\ \emph {et~al.}(2020)\citenamefont {Brandt},
  \citenamefont {Hiekkam{\"a}ki}, \citenamefont {Bouchard}, \citenamefont
  {Huber},\ and\ \citenamefont {Fickler}}]{Brandt2020}%
  \BibitemOpen
  \bibfield  {author} {\bibinfo {author} {\bibfnamefont {F.}~\bibnamefont
  {Brandt}}, \bibinfo {author} {\bibfnamefont {M.}~\bibnamefont
  {Hiekkam{\"a}ki}}, \bibinfo {author} {\bibfnamefont {F.}~\bibnamefont
  {Bouchard}}, \bibinfo {author} {\bibfnamefont {M.}~\bibnamefont {Huber}}, \
  and\ \bibinfo {author} {\bibfnamefont {R.}~\bibnamefont {Fickler}},\
  }\href@noop {} {\bibfield  {journal} {\bibinfo  {journal} {Optica}\ }\textbf
  {\bibinfo {volume} {7}},\ \bibinfo {pages} {98} (\bibinfo {year}
  {2020})}\BibitemShut {NoStop}%
\bibitem [{\citenamefont {Lu}\ \emph {et~al.}(2020)\citenamefont {Lu},
  \citenamefont {Simmerman}, \citenamefont {Lougovski}, \citenamefont
  {Weiner},\ and\ \citenamefont {Lukens}}]{Lu2020}%
  \BibitemOpen
  \bibfield  {author} {\bibinfo {author} {\bibfnamefont {H.-H.}\ \bibnamefont
  {Lu}}, \bibinfo {author} {\bibfnamefont {E.~M.}\ \bibnamefont {Simmerman}},
  \bibinfo {author} {\bibfnamefont {P.}~\bibnamefont {Lougovski}}, \bibinfo
  {author} {\bibfnamefont {A.~M.}\ \bibnamefont {Weiner}}, \ and\ \bibinfo
  {author} {\bibfnamefont {J.~M.}\ \bibnamefont {Lukens}},\ }\href@noop {}
  {\bibfield  {journal} {\bibinfo  {journal} {Phys. Rev. Lett.}\ }\textbf
  {\bibinfo {volume} {125}},\ \bibinfo {pages} {120503} (\bibinfo {year}
  {2020})}\BibitemShut {NoStop}%
\bibitem [{\citenamefont {Martin}\ \emph {et~al.}(2017)\citenamefont {Martin},
  \citenamefont {Guerreiro}, \citenamefont {Tiranov}, \citenamefont
  {Designolle}, \citenamefont {Fr{\"o}wis}, \citenamefont {Brunner},
  \citenamefont {Huber},\ and\ \citenamefont {Gisin}}]{Martin2017}%
  \BibitemOpen
  \bibfield  {author} {\bibinfo {author} {\bibfnamefont {A.}~\bibnamefont
  {Martin}}, \bibinfo {author} {\bibfnamefont {T.}~\bibnamefont {Guerreiro}},
  \bibinfo {author} {\bibfnamefont {A.}~\bibnamefont {Tiranov}}, \bibinfo
  {author} {\bibfnamefont {S.}~\bibnamefont {Designolle}}, \bibinfo {author}
  {\bibfnamefont {F.}~\bibnamefont {Fr{\"o}wis}}, \bibinfo {author}
  {\bibfnamefont {N.}~\bibnamefont {Brunner}}, \bibinfo {author} {\bibfnamefont
  {M.}~\bibnamefont {Huber}}, \ and\ \bibinfo {author} {\bibfnamefont
  {N.}~\bibnamefont {Gisin}},\ }\href@noop {} {\bibfield  {journal} {\bibinfo
  {journal} {Phys. Rev. Lett.}\ }\textbf {\bibinfo {volume} {118}},\ \bibinfo
  {pages} {110501} (\bibinfo {year} {2017})}\BibitemShut {NoStop}%
\bibitem [{\citenamefont {Ikuta}\ and\ \citenamefont
  {Takesue}(2017)}]{Ikuta2017}%
  \BibitemOpen
  \bibfield  {author} {\bibinfo {author} {\bibfnamefont {T.}~\bibnamefont
  {Ikuta}}\ and\ \bibinfo {author} {\bibfnamefont {H.}~\bibnamefont
  {Takesue}},\ }\href@noop {} {\bibfield  {journal} {\bibinfo  {journal} {New
  J. Phys.}\ }\textbf {\bibinfo {volume} {19}},\ \bibinfo {pages} {013039}
  (\bibinfo {year} {2017})}\BibitemShut {NoStop}%
\bibitem [{\citenamefont {Moody}\ \emph {et~al.}(2020)\citenamefont {Moody},
  \citenamefont {Chang}, \citenamefont {Steiner},\ and\ \citenamefont
  {Bowers}}]{Moody2020}%
  \BibitemOpen
  \bibfield  {author} {\bibinfo {author} {\bibfnamefont {G.}~\bibnamefont
  {Moody}}, \bibinfo {author} {\bibfnamefont {L.}~\bibnamefont {Chang}},
  \bibinfo {author} {\bibfnamefont {T.~J.}\ \bibnamefont {Steiner}}, \ and\
  \bibinfo {author} {\bibfnamefont {J.~E.}\ \bibnamefont {Bowers}},\
  }\href@noop {} {\bibfield  {journal} {\bibinfo  {journal} {AVS Quantum Sci.}\
  }\textbf {\bibinfo {volume} {2}},\ \bibinfo {pages} {041702} (\bibinfo {year}
  {2020})}\BibitemShut {NoStop}%
\bibitem [{\citenamefont {Yang}\ \emph {et~al.}(2021)\citenamefont {Yang},
  \citenamefont {Jahanbozorgi}, \citenamefont {Jeong}, \citenamefont {Sun},
  \citenamefont {Pfister}, \citenamefont {Lee},\ and\ \citenamefont
  {Yi}}]{Yang2021}%
  \BibitemOpen
  \bibfield  {author} {\bibinfo {author} {\bibfnamefont {Z.}~\bibnamefont
  {Yang}}, \bibinfo {author} {\bibfnamefont {M.}~\bibnamefont {Jahanbozorgi}},
  \bibinfo {author} {\bibfnamefont {D.}~\bibnamefont {Jeong}}, \bibinfo
  {author} {\bibfnamefont {S.}~\bibnamefont {Sun}}, \bibinfo {author}
  {\bibfnamefont {O.}~\bibnamefont {Pfister}}, \bibinfo {author} {\bibfnamefont
  {H.}~\bibnamefont {Lee}}, \ and\ \bibinfo {author} {\bibfnamefont
  {X.}~\bibnamefont {Yi}},\ }\href {\doibase 10.1038/s41467-021-25054-z}
  {\bibfield  {journal} {\bibinfo  {journal} {Nature Communications}\ }\textbf
  {\bibinfo {volume} {12}} (\bibinfo {year} {2021}),\
  10.1038/s41467-021-25054-z}\BibitemShut {NoStop}%
\bibitem [{\citenamefont {Elshaari}\ \emph {et~al.}(2020)\citenamefont
  {Elshaari}, \citenamefont {Pernice}, \citenamefont {Srinivasan},
  \citenamefont {Benson},\ and\ \citenamefont {Zwiller}}]{Elshaari2020}%
  \BibitemOpen
  \bibfield  {author} {\bibinfo {author} {\bibfnamefont {A.~W.}\ \bibnamefont
  {Elshaari}}, \bibinfo {author} {\bibfnamefont {W.}~\bibnamefont {Pernice}},
  \bibinfo {author} {\bibfnamefont {K.}~\bibnamefont {Srinivasan}}, \bibinfo
  {author} {\bibfnamefont {O.}~\bibnamefont {Benson}}, \ and\ \bibinfo {author}
  {\bibfnamefont {V.}~\bibnamefont {Zwiller}},\ }\href@noop {} {\bibfield
  {journal} {\bibinfo  {journal} {Nat. Photonics}\ }\textbf {\bibinfo {volume}
  {14}},\ \bibinfo {pages} {285} (\bibinfo {year} {2020})}\BibitemShut
  {NoStop}%
\bibitem [{\citenamefont {{Lu}}\ \emph {et~al.}(2019)\citenamefont {{Lu}},
  \citenamefont {{Weiner}}, \citenamefont {{Lougovski}},\ and\ \citenamefont
  {{Lukens}}}]{Lu2019c}%
  \BibitemOpen
  \bibfield  {author} {\bibinfo {author} {\bibfnamefont {H.-H.}\ \bibnamefont
  {{Lu}}}, \bibinfo {author} {\bibfnamefont {A.~M.}\ \bibnamefont {{Weiner}}},
  \bibinfo {author} {\bibfnamefont {P.}~\bibnamefont {{Lougovski}}}, \ and\
  \bibinfo {author} {\bibfnamefont {J.~M.}\ \bibnamefont {{Lukens}}},\
  }\href@noop {} {\bibfield  {journal} {\bibinfo  {journal} {IEEE Photon.
  Technol. Lett.}\ }\textbf {\bibinfo {volume} {31}},\ \bibinfo {pages} {1858}
  (\bibinfo {year} {2019})}\BibitemShut {NoStop}%
\bibitem [{\citenamefont {Lukens}\ \emph
  {et~al.}(2020{\natexlab{b}})\citenamefont {Lukens}, \citenamefont {Lu},
  \citenamefont {Qi}, \citenamefont {Lougovski}, \citenamefont {Weiner},\ and\
  \citenamefont {Williams}}]{Lukens2020a}%
  \BibitemOpen
  \bibfield  {author} {\bibinfo {author} {\bibfnamefont {J.~M.}\ \bibnamefont
  {Lukens}}, \bibinfo {author} {\bibfnamefont {H.-H.}\ \bibnamefont {Lu}},
  \bibinfo {author} {\bibfnamefont {B.}~\bibnamefont {Qi}}, \bibinfo {author}
  {\bibfnamefont {P.}~\bibnamefont {Lougovski}}, \bibinfo {author}
  {\bibfnamefont {A.~M.}\ \bibnamefont {Weiner}}, \ and\ \bibinfo {author}
  {\bibfnamefont {B.~P.}\ \bibnamefont {Williams}},\ }\href@noop {} {\bibfield
  {journal} {\bibinfo  {journal} {J. Light. Technol.}\ }\textbf {\bibinfo
  {volume} {38}},\ \bibinfo {pages} {1678} (\bibinfo {year}
  {2020}{\natexlab{b}})}\BibitemShut {NoStop}%
\bibitem [{\citenamefont {Liu}\ \emph {et~al.}(2020)\citenamefont {Liu},
  \citenamefont {Lucas}, \citenamefont {Raja}, \citenamefont {He},
  \citenamefont {Riemensberger}, \citenamefont {Wang}, \citenamefont {Karpov},
  \citenamefont {Guo}, \citenamefont {Bouchand},\ and\ \citenamefont
  {Kippenberg}}]{liu2020photonic}%
  \BibitemOpen
  \bibfield  {author} {\bibinfo {author} {\bibfnamefont {J.}~\bibnamefont
  {Liu}}, \bibinfo {author} {\bibfnamefont {E.}~\bibnamefont {Lucas}}, \bibinfo
  {author} {\bibfnamefont {A.~S.}\ \bibnamefont {Raja}}, \bibinfo {author}
  {\bibfnamefont {J.}~\bibnamefont {He}}, \bibinfo {author} {\bibfnamefont
  {J.}~\bibnamefont {Riemensberger}}, \bibinfo {author} {\bibfnamefont {R.~N.}\
  \bibnamefont {Wang}}, \bibinfo {author} {\bibfnamefont {M.}~\bibnamefont
  {Karpov}}, \bibinfo {author} {\bibfnamefont {H.}~\bibnamefont {Guo}},
  \bibinfo {author} {\bibfnamefont {R.}~\bibnamefont {Bouchand}}, \ and\
  \bibinfo {author} {\bibfnamefont {T.~J.}\ \bibnamefont {Kippenberg}},\
  }\href@noop {} {\bibfield  {journal} {\bibinfo  {journal} {Nat. Photonics}\
  }\textbf {\bibinfo {volume} {14}},\ \bibinfo {pages} {486} (\bibinfo {year}
  {2020})}\BibitemShut {NoStop}%
\bibitem [{\citenamefont {Robert}\ and\ \citenamefont
  {Casella}(1999)}]{Robert1999}%
  \BibitemOpen
  \bibfield  {author} {\bibinfo {author} {\bibfnamefont {C.~P.}\ \bibnamefont
  {Robert}}\ and\ \bibinfo {author} {\bibfnamefont {G.}~\bibnamefont
  {Casella}},\ }\href@noop {} {\emph {\bibinfo {title} {Monte Carlo Statistical
  Methods}}}\ (\bibinfo  {publisher} {Springer},\ \bibinfo {address} {New
  York},\ \bibinfo {year} {1999})\BibitemShut {NoStop}%
\bibitem [{\citenamefont {Peres}(1996)}]{Peres1996}%
  \BibitemOpen
  \bibfield  {author} {\bibinfo {author} {\bibfnamefont {A.}~\bibnamefont
  {Peres}},\ }\href@noop {} {\bibfield  {journal} {\bibinfo  {journal} {Phys.
  Rev. Lett.}\ }\textbf {\bibinfo {volume} {77}},\ \bibinfo {pages} {1413}
  (\bibinfo {year} {1996})}\BibitemShut {NoStop}%
\bibitem [{\citenamefont {Vidal}\ and\ \citenamefont
  {Werner}(2002)}]{Vidal2002}%
  \BibitemOpen
  \bibfield  {author} {\bibinfo {author} {\bibfnamefont {G.}~\bibnamefont
  {Vidal}}\ and\ \bibinfo {author} {\bibfnamefont {R.~F.}\ \bibnamefont
  {Werner}},\ }\href@noop {} {\bibfield  {journal} {\bibinfo  {journal} {Phys.
  Rev. A}\ }\textbf {\bibinfo {volume} {65}},\ \bibinfo {pages} {032314}
  (\bibinfo {year} {2002})}\BibitemShut {NoStop}%
\bibitem [{\citenamefont {Imany}\ \emph {et~al.}(2020)\citenamefont {Imany},
  \citenamefont {Lingaraju}, \citenamefont {Alshaykh}, \citenamefont {Leaird},\
  and\ \citenamefont {Weiner}}]{Imany2020}%
  \BibitemOpen
  \bibfield  {author} {\bibinfo {author} {\bibfnamefont {P.}~\bibnamefont
  {Imany}}, \bibinfo {author} {\bibfnamefont {N.~B.}\ \bibnamefont
  {Lingaraju}}, \bibinfo {author} {\bibfnamefont {M.~S.}\ \bibnamefont
  {Alshaykh}}, \bibinfo {author} {\bibfnamefont {D.~E.}\ \bibnamefont
  {Leaird}}, \ and\ \bibinfo {author} {\bibfnamefont {A.~M.}\ \bibnamefont
  {Weiner}},\ }\href {\doibase 10.1126/sciadv.aba8066} {\bibfield  {journal}
  {\bibinfo  {journal} {Sci. Adv.}\ }\textbf {\bibinfo {volume} {6}},\ \bibinfo
  {pages} {eaba8066} (\bibinfo {year} {2020})}\BibitemShut {NoStop}%
\bibitem [{\citenamefont {Gross}\ \emph {et~al.}(2010)\citenamefont {Gross},
  \citenamefont {Liu}, \citenamefont {Flammia}, \citenamefont {Becker},\ and\
  \citenamefont {Eisert}}]{Gross2010}%
  \BibitemOpen
  \bibfield  {author} {\bibinfo {author} {\bibfnamefont {D.}~\bibnamefont
  {Gross}}, \bibinfo {author} {\bibfnamefont {Y.-K.}\ \bibnamefont {Liu}},
  \bibinfo {author} {\bibfnamefont {S.~T.}\ \bibnamefont {Flammia}}, \bibinfo
  {author} {\bibfnamefont {S.}~\bibnamefont {Becker}}, \ and\ \bibinfo {author}
  {\bibfnamefont {J.}~\bibnamefont {Eisert}},\ }\href {\doibase
  10.1103/PhysRevLett.105.150401} {\bibfield  {journal} {\bibinfo  {journal}
  {Phys. Rev. Lett.}\ }\textbf {\bibinfo {volume} {105}},\ \bibinfo {pages}
  {150401} (\bibinfo {year} {2010})}\BibitemShut {NoStop}%
\bibitem [{\citenamefont {Brydges}\ \emph {et~al.}(2019)\citenamefont
  {Brydges}, \citenamefont {Elben}, \citenamefont {Jurcevic}, \citenamefont
  {Vermersch}, \citenamefont {Maier}, \citenamefont {Lanyon}, \citenamefont
  {Zoller}, \citenamefont {Blatt},\ and\ \citenamefont {Roos}}]{Brydges2019}%
  \BibitemOpen
  \bibfield  {author} {\bibinfo {author} {\bibfnamefont {T.}~\bibnamefont
  {Brydges}}, \bibinfo {author} {\bibfnamefont {A.}~\bibnamefont {Elben}},
  \bibinfo {author} {\bibfnamefont {P.}~\bibnamefont {Jurcevic}}, \bibinfo
  {author} {\bibfnamefont {B.}~\bibnamefont {Vermersch}}, \bibinfo {author}
  {\bibfnamefont {C.}~\bibnamefont {Maier}}, \bibinfo {author} {\bibfnamefont
  {B.~P.}\ \bibnamefont {Lanyon}}, \bibinfo {author} {\bibfnamefont
  {P.}~\bibnamefont {Zoller}}, \bibinfo {author} {\bibfnamefont
  {R.}~\bibnamefont {Blatt}}, \ and\ \bibinfo {author} {\bibfnamefont {C.~F.}\
  \bibnamefont {Roos}},\ }\href {\doibase 10.1126/science.aau4963} {\bibfield
  {journal} {\bibinfo  {journal} {Science}\ }\textbf {\bibinfo {volume}
  {364}},\ \bibinfo {pages} {260} (\bibinfo {year} {2019})}\BibitemShut
  {NoStop}%
\bibitem [{\citenamefont {Huang}\ \emph {et~al.}(2020)\citenamefont {Huang},
  \citenamefont {Kueng},\ and\ \citenamefont {Preskill}}]{Huang2020}%
  \BibitemOpen
  \bibfield  {author} {\bibinfo {author} {\bibfnamefont {H.-Y.}\ \bibnamefont
  {Huang}}, \bibinfo {author} {\bibfnamefont {R.}~\bibnamefont {Kueng}}, \ and\
  \bibinfo {author} {\bibfnamefont {J.}~\bibnamefont {Preskill}},\ }\href@noop
  {} {\bibfield  {journal} {\bibinfo  {journal} {Nat. Phys.}\ }\textbf
  {\bibinfo {volume} {16}},\ \bibinfo {pages} {1050} (\bibinfo {year}
  {2020})}\BibitemShut {NoStop}%
\bibitem [{\citenamefont {Liu}\ \emph {et~al.}(2021)\citenamefont {Liu},
  \citenamefont {Huang}, \citenamefont {Wang}, \citenamefont {He},
  \citenamefont {Raja}, \citenamefont {Liu}, \citenamefont {Engelsen},\ and\
  \citenamefont {Kippenberg}}]{liu2021high}%
  \BibitemOpen
  \bibfield  {author} {\bibinfo {author} {\bibfnamefont {J.}~\bibnamefont
  {Liu}}, \bibinfo {author} {\bibfnamefont {G.}~\bibnamefont {Huang}}, \bibinfo
  {author} {\bibfnamefont {R.~N.}\ \bibnamefont {Wang}}, \bibinfo {author}
  {\bibfnamefont {J.}~\bibnamefont {He}}, \bibinfo {author} {\bibfnamefont
  {A.~S.}\ \bibnamefont {Raja}}, \bibinfo {author} {\bibfnamefont
  {T.}~\bibnamefont {Liu}}, \bibinfo {author} {\bibfnamefont {N.~J.}\
  \bibnamefont {Engelsen}}, \ and\ \bibinfo {author} {\bibfnamefont {T.~J.}\
  \bibnamefont {Kippenberg}},\ }\href@noop {} {\bibfield  {journal} {\bibinfo
  {journal} {Nat. Commun.}\ }\textbf {\bibinfo {volume} {12}},\ \bibinfo
  {pages} {1} (\bibinfo {year} {2021})}\BibitemShut {NoStop}%
\bibitem [{\citenamefont {Liu}\ \emph {et~al.}(2018)\citenamefont {Liu},
  \citenamefont {Raja}, \citenamefont {Pfeiffer}, \citenamefont {Herkommer},
  \citenamefont {Guo}, \citenamefont {Zervas}, \citenamefont {Geiselmann},\
  and\ \citenamefont {Kippenberg}}]{liu2018double}%
  \BibitemOpen
  \bibfield  {author} {\bibinfo {author} {\bibfnamefont {J.}~\bibnamefont
  {Liu}}, \bibinfo {author} {\bibfnamefont {A.~S.}\ \bibnamefont {Raja}},
  \bibinfo {author} {\bibfnamefont {M.~H.}\ \bibnamefont {Pfeiffer}}, \bibinfo
  {author} {\bibfnamefont {C.}~\bibnamefont {Herkommer}}, \bibinfo {author}
  {\bibfnamefont {H.}~\bibnamefont {Guo}}, \bibinfo {author} {\bibfnamefont
  {M.}~\bibnamefont {Zervas}}, \bibinfo {author} {\bibfnamefont
  {M.}~\bibnamefont {Geiselmann}}, \ and\ \bibinfo {author} {\bibfnamefont
  {T.~J.}\ \bibnamefont {Kippenberg}},\ }\href@noop {} {\bibfield  {journal}
  {\bibinfo  {journal} {Opt. Lett.}\ }\textbf {\bibinfo {volume} {43}},\
  \bibinfo {pages} {3200} (\bibinfo {year} {2018})}\BibitemShut {NoStop}%
\bibitem [{\citenamefont {Myilswamy}\ \emph {et~al.}(2021)\citenamefont
  {Myilswamy}, \citenamefont {Alshaykh}, \citenamefont {Lu}, \citenamefont
  {Liu}, \citenamefont {Leaird}, \citenamefont {Kippenberg},\ and\
  \citenamefont {Weiner}}]{KarthikCLEO}%
  \BibitemOpen
  \bibfield  {author} {\bibinfo {author} {\bibfnamefont {K.~V.}\ \bibnamefont
  {Myilswamy}}, \bibinfo {author} {\bibfnamefont {M.~S.}\ \bibnamefont
  {Alshaykh}}, \bibinfo {author} {\bibfnamefont {H.-H.}\ \bibnamefont {Lu}},
  \bibinfo {author} {\bibfnamefont {J.}~\bibnamefont {Liu}}, \bibinfo {author}
  {\bibfnamefont {D.~E.}\ \bibnamefont {Leaird}}, \bibinfo {author}
  {\bibfnamefont {T.~J.}\ \bibnamefont {Kippenberg}}, \ and\ \bibinfo {author}
  {\bibfnamefont {A.~M.}\ \bibnamefont {Weiner}},\ }in\ \href@noop {} {\emph
  {\bibinfo {booktitle} {CLEO: 2021}}}\ (\bibinfo  {publisher} {Optical Society
  of America},\ \bibinfo {year} {2021})\ p.\ \bibinfo {pages}
  {JM3F.5}\BibitemShut {NoStop}%
\bibitem [{\citenamefont {Lu}\ \emph {et~al.}(2018{\natexlab{b}})\citenamefont
  {Lu}, \citenamefont {Lukens}, \citenamefont {Peters}, \citenamefont {Odele},
  \citenamefont {Leaird}, \citenamefont {Weiner},\ and\ \citenamefont
  {Lougovski}}]{Lu2018a}%
  \BibitemOpen
  \bibfield  {author} {\bibinfo {author} {\bibfnamefont {H.-H.}\ \bibnamefont
  {Lu}}, \bibinfo {author} {\bibfnamefont {J.~M.}\ \bibnamefont {Lukens}},
  \bibinfo {author} {\bibfnamefont {N.~A.}\ \bibnamefont {Peters}}, \bibinfo
  {author} {\bibfnamefont {O.~D.}\ \bibnamefont {Odele}}, \bibinfo {author}
  {\bibfnamefont {D.~E.}\ \bibnamefont {Leaird}}, \bibinfo {author}
  {\bibfnamefont {A.~M.}\ \bibnamefont {Weiner}}, \ and\ \bibinfo {author}
  {\bibfnamefont {P.}~\bibnamefont {Lougovski}},\ }\href@noop {} {\bibfield
  {journal} {\bibinfo  {journal} {Phys. Rev. Lett.}\ }\textbf {\bibinfo
  {volume} {120}},\ \bibinfo {pages} {030502} (\bibinfo {year}
  {2018}{\natexlab{b}})}\BibitemShut {NoStop}%
\bibitem [{\citenamefont {Williams}\ and\ \citenamefont
  {Lougovski}(2017)}]{Williams2017}%
  \BibitemOpen
  \bibfield  {author} {\bibinfo {author} {\bibfnamefont {B.~P.}\ \bibnamefont
  {Williams}}\ and\ \bibinfo {author} {\bibfnamefont {P.}~\bibnamefont
  {Lougovski}},\ }\href {http://stacks.iop.org/1367-2630/19/i=4/a=043003}
  {\bibfield  {journal} {\bibinfo  {journal} {New J. Phys.}\ }\textbf {\bibinfo
  {volume} {19}},\ \bibinfo {pages} {043003} (\bibinfo {year}
  {2017})}\BibitemShut {NoStop}%
\bibitem [{\citenamefont {Lu}\ \emph {et~al.}(2019)\citenamefont {Lu},
  \citenamefont {Lukens}, \citenamefont {Williams}, \citenamefont {Imany},
  \citenamefont {Peters}, \citenamefont {Weiner},\ and\ \citenamefont
  {Lougovski}}]{Lu2019a}%
  \BibitemOpen
  \bibfield  {author} {\bibinfo {author} {\bibfnamefont {H.-H.}\ \bibnamefont
  {Lu}}, \bibinfo {author} {\bibfnamefont {J.~M.}\ \bibnamefont {Lukens}},
  \bibinfo {author} {\bibfnamefont {B.~P.}\ \bibnamefont {Williams}}, \bibinfo
  {author} {\bibfnamefont {P.}~\bibnamefont {Imany}}, \bibinfo {author}
  {\bibfnamefont {N.~A.}\ \bibnamefont {Peters}}, \bibinfo {author}
  {\bibfnamefont {A.~M.}\ \bibnamefont {Weiner}}, \ and\ \bibinfo {author}
  {\bibfnamefont {P.}~\bibnamefont {Lougovski}},\ }\href@noop {} {\bibfield
  {journal} {\bibinfo  {journal} {npj Quantum Inf.}\ }\textbf {\bibinfo
  {volume} {5}},\ \bibinfo {pages} {24} (\bibinfo {year} {2019})}\BibitemShut
  {NoStop}%
\bibitem [{\citenamefont {Simmerman}\ \emph {et~al.}(2020)\citenamefont
  {Simmerman}, \citenamefont {Lu}, \citenamefont {Weiner},\ and\ \citenamefont
  {Lukens}}]{Simmerman2020}%
  \BibitemOpen
  \bibfield  {author} {\bibinfo {author} {\bibfnamefont {E.~M.}\ \bibnamefont
  {Simmerman}}, \bibinfo {author} {\bibfnamefont {H.-H.}\ \bibnamefont {Lu}},
  \bibinfo {author} {\bibfnamefont {A.~M.}\ \bibnamefont {Weiner}}, \ and\
  \bibinfo {author} {\bibfnamefont {J.~M.}\ \bibnamefont {Lukens}},\ }\href
  {\doibase 10.1364/OL.392694} {\bibfield  {journal} {\bibinfo  {journal} {Opt.
  Lett.}\ }\textbf {\bibinfo {volume} {45}},\ \bibinfo {pages} {2886} (\bibinfo
  {year} {2020})}\BibitemShut {NoStop}%
\bibitem [{\citenamefont {Sommers}\ and\ \citenamefont
  {\.{Z}yczkowski}(2003)}]{Sommers2003}%
  \BibitemOpen
  \bibfield  {author} {\bibinfo {author} {\bibfnamefont {H.-J.}\ \bibnamefont
  {Sommers}}\ and\ \bibinfo {author} {\bibfnamefont {K.}~\bibnamefont
  {\.{Z}yczkowski}},\ }\href {\doibase 10.1088/0305-4470/36/39/308} {\bibfield
  {journal} {\bibinfo  {journal} {J. Phys. A: Math. Gen.}\ }\textbf {\bibinfo
  {volume} {36}},\ \bibinfo {pages} {10083} (\bibinfo {year}
  {2003})}\BibitemShut {NoStop}%
\bibitem [{\citenamefont {Osipov}\ \emph {et~al.}(2010)\citenamefont {Osipov},
  \citenamefont {Sommers},\ and\ \citenamefont
  {{\.{Z}}yczkowski}}]{Osipov2010}%
  \BibitemOpen
  \bibfield  {author} {\bibinfo {author} {\bibfnamefont {V.~A.}\ \bibnamefont
  {Osipov}}, \bibinfo {author} {\bibfnamefont {H.-J.}\ \bibnamefont {Sommers}},
  \ and\ \bibinfo {author} {\bibfnamefont {K.}~\bibnamefont
  {{\.{Z}}yczkowski}},\ }\href {\doibase 10.1088/1751-8113/43/5/055302}
  {\bibfield  {journal} {\bibinfo  {journal} {J. Phys. A: Math. Theor.}\
  }\textbf {\bibinfo {volume} {43}},\ \bibinfo {pages} {055302} (\bibinfo
  {year} {2010})}\BibitemShut {NoStop}%
\bibitem [{\citenamefont {Mezzadri}(2007)}]{Mezzadri2007}%
  \BibitemOpen
  \bibfield  {author} {\bibinfo {author} {\bibfnamefont {F.}~\bibnamefont
  {Mezzadri}},\ }\href@noop {} {\bibfield  {journal} {\bibinfo  {journal} {Not.
  Am. Math. Soc.}\ }\textbf {\bibinfo {volume} {54}},\ \bibinfo {pages} {592}
  (\bibinfo {year} {2007})}\BibitemShut {NoStop}%
\bibitem [{\citenamefont {MacKay}(2003)}]{MacKay2003}%
  \BibitemOpen
  \bibfield  {author} {\bibinfo {author} {\bibfnamefont {D.~J.~C.}\
  \bibnamefont {MacKay}},\ }\href@noop {} {\emph {\bibinfo {title}
  {{Information Theory, Inference, and Learning Algorithms}}}}\ (\bibinfo
  {publisher} {Cambridge University Press},\ \bibinfo {address} {Cambridge,
  UK},\ \bibinfo {year} {2003})\BibitemShut {NoStop}%
\bibitem [{\citenamefont {Cotter}\ \emph {et~al.}(2013)\citenamefont {Cotter},
  \citenamefont {Roberts}, \citenamefont {Stuart},\ and\ \citenamefont
  {White}}]{Cotter2013}%
  \BibitemOpen
  \bibfield  {author} {\bibinfo {author} {\bibfnamefont {S.~L.}\ \bibnamefont
  {Cotter}}, \bibinfo {author} {\bibfnamefont {G.~O.}\ \bibnamefont {Roberts}},
  \bibinfo {author} {\bibfnamefont {A.~M.}\ \bibnamefont {Stuart}}, \ and\
  \bibinfo {author} {\bibfnamefont {D.}~\bibnamefont {White}},\ }\href
  {\doibase 10.1214/13-STS421} {\bibfield  {journal} {\bibinfo  {journal}
  {Statist. Sci.}\ }\textbf {\bibinfo {volume} {28}},\ \bibinfo {pages} {424}
  (\bibinfo {year} {2013})}\BibitemShut {NoStop}%
\bibitem [{\citenamefont {Jacob}\ \emph {et~al.}(2020)\citenamefont {Jacob},
  \citenamefont {O'Leary},\ and\ \citenamefont {Atchad\'{e}}}]{Jacob2020}%
  \BibitemOpen
  \bibfield  {author} {\bibinfo {author} {\bibfnamefont {P.~E.}\ \bibnamefont
  {Jacob}}, \bibinfo {author} {\bibfnamefont {J.}~\bibnamefont {O'Leary}}, \
  and\ \bibinfo {author} {\bibfnamefont {Y.~F.}\ \bibnamefont {Atchad\'{e}}},\
  }\href {\doibase https://doi.org/10.1111/rssb.12336} {\bibfield  {journal}
  {\bibinfo  {journal} {J. R. Statist. Soc. B}\ }\textbf {\bibinfo {volume}
  {82}},\ \bibinfo {pages} {543} (\bibinfo {year} {2020})}\BibitemShut
  {NoStop}%
\bibitem [{\citenamefont {Wootters}\ and\ \citenamefont
  {Fields}(1989)}]{Wootters1989}%
  \BibitemOpen
  \bibfield  {author} {\bibinfo {author} {\bibfnamefont {W.~K.}\ \bibnamefont
  {Wootters}}\ and\ \bibinfo {author} {\bibfnamefont {B.~D.}\ \bibnamefont
  {Fields}},\ }\href {\doibase http://dx.doi.org/10.1016/0003-4916(89)90322-9}
  {\bibfield  {journal} {\bibinfo  {journal} {Ann. Phys.}\ }\textbf {\bibinfo
  {volume} {191}},\ \bibinfo {pages} {363} (\bibinfo {year}
  {1989})}\BibitemShut {NoStop}%
\end{thebibliography}
\end{document}